\documentclass[12pt,preprint]{aastex}
\shorttitle{Second order self-gravity calculation in polar coordinates}
\shortauthors{Wang et al.}
\usepackage{bm}
\usepackage{sansmath}
\usepackage{amsmath}
\usepackage{mathrsfs}
\usepackage{epstopdf}
\usepackage[breaklinks,colorlinks,urlcolor=blue,citecolor=blue,linkcolor=blue]{hyperref}
\usepackage{booktabs}
\newcommand{\vect}[1]{\boldsymbol{#1}}
\newcommand{\hhw}[1]{{\color{black} #1}}
\newcommand{\rt}[1]{{\color{black} #1}}
\defcitealias{Bar08}{BM08}

\begin{document}

\title{Self-Gravitational Force Calculation of Second Order Accuracy for Infinitesimally Thin Gaseous Disks in Polar Coordinates}

\author{Hsiang-Hsu Wang\altaffilmark{1}, David C. C. Yen \altaffilmark{2}, Ronald E. Taam\altaffilmark{1,3}}
\email{yen@math.fju.edu.tw}
\altaffiltext{1}{Institute of Astronomy and Astrophysics, Academia Sinica, P.O. Box 23-141, Taipei 10617, Taiwan, R.O.C.}
\altaffiltext{2}{Department of Mathematics, Fu Jen Catholic University, New Taipei City, Taiwan.}
\altaffiltext{3}{Department of Physics and Astronomy, Northwestern University, 2131 Tech Drive, Evanston, IL 60208, USA}

\begin{abstract}
Investigating the evolution of disk galaxies and the dynamics of proto-stellar disks can involve the 
use of both 
a hydrodynamical and a Poisson solver. These systems are usually approximated as infinitesimally thin 
disks using two-dimensional Cartesian or polar coordinates. In Cartesian coordinates, the calculations of 
the hydrodynamics and self-gravitational forces are relatively straightforward for attaining 
second order accuracy. However, in polar coordinates, a second order calculation of self-gravitational 
forces is required for matching the second order accuracy of hydrodynamical schemes. We present 
a direct algorithm for calculating self-gravitational forces with second order accuracy without artificial 
boundary conditions. The Poisson integral in polar coordinates is expressed in a convolution form and the 
corresponding numerical complexity is nearly linear using a fast Fourier transform.  Examples with 
analytic solutions are used to verify that the truncated error of this algorithm is of second order.  The 
kernel integral around the singularity is applied to modify the particle method.  The use of a 
softening length is avoided and the accuracy of the particle method is significantly improved. 
\end{abstract}


\section{Introduction}
Thin disks are common in the Universe as a result of the conservation of angular momentum and efficient 
radiative cooling. The existence of 
central starburst rings \citep{Lin13,Seo14}, bright and young stars formed along spiral arms \citep{Elm14}, 
and substructures associated with bars and spirals \citep{Kim12,Lee12,Lee14} indicate that the self-gravity 
of gas is important to the evolution of disk galaxies.  The formation of planets in the early phase of 
proto-stellar disks indicates that self-gravity of gaseous disks plays a role in shaping planetary 
systems \citep{Zha08,Inu10,Zha14}. As a first approximation, these thin disks are usually studied using 
two-dimensional hydrodynamical simulations coupled with a Poisson solver. 

For a given mass distribution, the calculation of self-gravitational forces is a fundamental, but a 
challenging aspect of computational astrophysics.  For three-dimensional grid-based codes, several 
techniques have been proposed to improve the accuracy and the performance of calculations.  Of the 
simplest ones is the use of the fast Fourier transform (FFT), which is suitable for both periodic and 
isolated boundary conditions \citep{Jam77}.  The FFT techniques are fast, accurate and are suitable for 
both three- and two-dimensional calculations.  The multigrid relaxation methods are fast, flexible and 
have been used extensively when mesh refinements are required \citep{Hoc88}.  However, the multigrid 
methods, which are by nature only for three-dimensional problems, cannot be reduced to two-dimensional 
calculations for an infinitesimally thin disk as discussed in this paper. 

Compared to those techniques well developed for Cartesian coordinates, the calculation of the 
self-gravitational force in cylindrical and polar coordinates still requires further study. 
\citet{Yen12} developed formulae for the calculation of these forces to 2nd-order accuracy for both 
Cartesian and polar coordinates.  In this description, the Poisson integral is written in 
convolution form and is of linear complexity if the FFT is used.  Unlike those Poisson integrals 
which are integrable in Cartesian coordinates, no closed forms were found for polar coordinates 
because the elliptic integral is involved.  Consequently, while the 2nd-order accuracy can be achieved in 
Cartesian coordinates, the calculations in polar coordinates suffer from the presence of a 
singularity in the kernel integral, reducing the order of convergence to nearly first order. Convolution 
expressions of the Poisson integrals in polar coordinates is also adopted in \citet[][hereafter BM08]{Bar08} 
\rt{in their two dimensional hydrodynamical calculations}. However, in their formulae, the use of a softening 
length\hhw{, though physically} \rt{motivated, inhibits pursuing higher order accuracy}.

In this work, we develop a simple, but effective algorithm that increases the accuracy of the 
self-gravitational calculations in polar coordinates to 2nd-order.  The proposed method retains a 
linear complexity since all the effort is directed to the preparation of accurate force kernels. 
The technique developed in this work is applied to improve the numerical accuracy of the particle 
method.  The use of a softening length is avoided and the force kernel integrals in the neighborhood 
of a singularity significantly reduce the numerical error of the particle method.

This paper is organized as follows.  The framework and assumptions adopted for this work are outlined 
in \S~2. We develop the mathematical notations and formulae for the calculation of the 
self-gravitational force to 2nd-order accuracy and the modified particle methods in both Cartesian and 
cylindrical coordinates, respectively in \S~3. The 2nd-order method described for Cartesian coordinates 
is used in \S~4, where two improvements for the evaluation of force kernels in cylindrical 
coordinates are elaborated.  In \S~5, detailed comparisons between the numerical results and analytic 
solutions are discussed.  We summarize our results and conclude in \S~6.  

\section{Framework and assumptions} 

The potential $\Phi$ for a given distribution of gaseous density $\rho$ in three-dimensional space
satisfies the Poisson equation  below:
\begin{equation}
\nabla^2 \Phi(\vect{x})=4\pi G \rho(\vect{x}),
\end{equation}
where $G$ is the gravitational constant and $\vect{x}$ denotes the position vector. Without loss of generality,
 we assume that $G=1$ throughout this work. By imposing the boundary condition,
\begin{equation}
\lim_{|\vect{x}|\rightarrow\infty}\Phi(\vect{x})=0,
\end{equation}
the gravitational potential $\Phi(\vect{x})$ can be cast in an integral form~\citep{Eva91, Bin08}:
\begin{equation}
\label{Eq:Phi_Int}
\Phi(x,y,z)=-\int\int\int \mathcal{K}(\bar{x}-x,\bar{y}-y,\bar{z}-z)\rho(\bar{x},\bar{y},\bar{z}){\rm d}\bar{x}{\rm d}\bar{y}{\rm d}\bar{z},
\end{equation}
where $(x,y,z)$ are Cartesian coordinates and $\mathcal{K}\equiv 1/\sqrt{x^2+y^2+z^2}$ is the kernel of the integral. 
In this paper, we restrict the discussion to the following expression for the density distribution:
\begin{equation}
\rho(\vect{x})=\sigma(x,y)\delta(z),
\end{equation}
where $\delta$ denotes the Dirac delta function and $\sigma(x,y)$ is the surface density defined as:
\begin{equation}
\label{Eq:sigma_def}
\sigma(x,y)=\int \rho(\vect{x}){\rm d}z.
\end{equation}
The integral form of the gravitational potential, i.e., Equation~(\ref{Eq:Phi_Int}), 
and the associated forces can be numerically evaluated through discretization in the 
computational domain.  \citet{Yen12} have shown that uniform discretization in Cartesian coordinates 
and radially logarithmic discretization in polar coordinates enable the self-gravitating forces 
to be expressed in a convolution form of a double summation. With the assumption that the density 
distribution is smooth, the linear approximation for the surface density in each cell increases 
the accuracy of numerical solution.  Using the convolution theorem \citep{Bra99}, a fast Fourier transform 
is applied to reduce the computational complexity from $O(N^4)$ to $O(N^2\log_2N)$, where $N$ is the number 
of zones in one direction.  This method is a direct calculation of the self-gravitational forces, which 
is second order accurate in Cartesian coordinates, without necessarily invoking the use of artificial 
boundary conditions. 

One of the major advantages of the convolution integral approach is that most of the calculational 
effort is passed to the preparation of the kernels.  That is, the more accurate the kernels, 
the more accurate the numerical solutions.  We note that those kernels used to achieve higher order accuracy 
need to be prepared only once for a fixed grid and stored in the computer memory at the beginning of 
each simulation.  In 
the following sections, we exploit the advantage of using the convolution integral and restrict the 
associated discussions to the plane of the disk, i.e., $\Phi(x,y,0)$. Simple but effective approaches are 
proposed to improve the numerical accuracy and the order of convergence. 

\section{A direct method of 2nd-order accuracy and a modified particle method}
In this section, we develop the mathematical notations that will be used throughout this work so that 
the material in this paper is self-contained.  The expressions of formulae with 2nd-order accuracy are 
first derived in Cartesian and polar coordinates for readability and completeness.  Based on the 
2nd-order method, approximations are adopted to reduce the computational cost further by concentrating the 
mass of one cell at the cell center.  The simplified scheme is a modified particle method.  The use 
of a softening length is avoided and the associated singularity problem is removed using the kernel 
integrals.  Without increasing the computational cost, the modified particle-based method significantly 
improves the accuracy of the numerical solutions. 

In the following, we discuss in detail the calculations of the self-gravitating forces in the 
$x$-direction for Cartesian coordinates and in the $r$-direction for polar coordinates.  In Appendix 
A, we provide the formulae for the calculations of the self-gravitational forces in the $y$-direction 
and $\phi$-direction.  The full expressions for the kernel integrals are also given.

\subsection{Cartesian coordinates}

\subsubsection{A direct method of 2nd-order accuracy}
Consider a calculational domain described by $D=[-M,M]\times[-M,M]$ for some number $M>0$, which 
is evenly subdivided with $N_{\rm d}$ intervals in the $x$- and $y$-directions, respectively. Given a 
positive number $N_{\rm d}$, we define $\Delta x=2M/N_{\rm d}$, $\Delta y=2M/N_{\rm d}$ as the cell size 
in each direction and $x_{i+1/2}=-M+i\Delta x$, $y_{j+1/2}=-M+j\Delta y$ as the cell boundaries, where 
$i,j=0,...,N_{\rm d}$. The domain of each cell is then defined to be $D_{i,j}\equiv [x_{i-1/2},x_{i+1/2}]
\times[y_{j-1/2},y_{j+1/2}]$ and the cell centers are $x_i=(x_{i-1/2}+x_{i+1/2})/2$, $y_j=(y_{j-1/2}+
y_{j+1/2})/2$, with $i,j=1,...,N_{\rm d}$. In total, the calculational domain is covered with 
$N_{\rm d}^2$ cells.

The forces in the $x$-direction ($F^x_{i,j}$) and $y$-directions ($F^y_{i,j}$) are defined at the 
center of cells and related to Equation~(\ref{Eq:Phi_Int}) through the following relations:
\begin{eqnarray}
\label{Eq:Fx_def}
F^x_{i,j}\equiv-\frac{\partial}{\partial x}\Phi(x_i,y_j,0)=\sum_{i'=1}^{N_{\rm d}}\sum_{j'=1}^{N_{\rm d}}\int\int_{D_{i',j'}}\frac{\partial}{\partial x}\mathcal{K}(\bar{x}-x_i,\bar{y}-y_j,0)\sigma(\bar{x},\bar{y}){\rm d}\bar{x}{\rm d}\bar{y}\\ 
F^y_{i,j}\equiv-\frac{\partial}{\partial y}\Phi(x_i,y_j,0)=\sum_{i'=1}^{N_{\rm d}}\sum_{j'=1}^{N_{\rm d}}\int\int_{D_{i',j'}}\frac{\partial}{\partial y}\mathcal{K}(\bar{x}-x_i,\bar{y}-y_j,0)\sigma(\bar{x},\bar{y}){\rm d}\bar{x}{\rm d}\bar{y}.
\end{eqnarray}
The surface density $\sigma$ in cells, appearing in Equations~(6) and (7) can be linearly 
approximated by
\begin{equation}
\label{Eq:sigma_approx}
\sigma(\bar{x},\bar{y})\approx\sigma_{i',j'}+\delta^x_{i',j'}(\bar{x}-x_{i'})+\delta^y_{i',j'}(\bar{y}-y_{j'})
\end{equation}
where $\sigma_{i',j'}$, $\delta^x_{i',j'}\equiv \partial \sigma(x_{i'},y_{j'})/\partial x$ and 
$\delta^y_{i',j'}\equiv \partial \sigma(x_{i'},y_{j'})/\partial y$ are constant in the cell $D_{i',j'}$. 
With the linear approximation in surface density, $F^x_{i,j}$ with second order of accuracy can be 
approximated by \citep{Yen12}:
\begin{equation}
\label{Eq:Fx_approx}
F^x_{i,j} \approx F^{x,0}_{i,j}+F^{x,x}_{i,j}+F^{x,y}_{i,j},
\end{equation}
where 
\begin{eqnarray}
F^{x,0}_{i,j}=\sum_{i'=1}^{N_{\rm d}}\sum_{j'=1}^{N_{\rm d}} \sigma_{i',j'} \mathcal{K}^{x,0}_{i-i',j-j'}, \\
F^{x,x}_{i,j}=\sum_{i'=1}^{N_{\rm d}}\sum_{j'=1}^{N_{\rm d}} \delta^x_{i',j'} \mathcal{K}^{x,x}_{i-i',j-j'}, \\
F^{x,y}_{i,j}=\sum_{i'=1}^{N_{\rm d}}\sum_{j'=1}^{N_{\rm d}} \delta^y_{i',j'} \mathcal{K}^{x,y}_{i-i',j-j'},
\end{eqnarray}
and 
\begin{eqnarray}
\mathcal{K}^{x,0}_{i-i',j-j'}\equiv\int\int_{D_{i',j'}} \frac{\bar{x}-x_i}{[(\bar{x}-x_i)^2+(\bar{y}-y_j)^2]^{3/2}}{\rm d}\bar{x}{\rm d}\bar{y}, \\
\mathcal{K}^{x,x}_{i-i',j-j'}\equiv\int\int_{D_{i',j'}} \frac{(\bar{x}-x_i)(\bar{x}-x_{i'})}{[(\bar{x}-x_i)^2+(\bar{y}-y_j)^2]^{3/2}}{\rm d}\bar{x}{\rm d}\bar{y}, \\
\mathcal{K}^{x,y}_{i-i',j-j'}\equiv\int\int_{D_{i',j'}} \frac{(\bar{x}-x_i)(\bar{y}-y_{j'})}{[(\bar{x}-x_i)^2+(\bar{y}-y_j)^2]^{3/2}}{\rm d}\bar{x}{\rm d}\bar{y}.
\end{eqnarray}
The first term in Equation~(\ref{Eq:Fx_approx}) is the contribution if the mass enclosed within one cell 
were uniformly distributed and provides an accuracy of first order.  On the other hand, the last 
two terms take into account the structure of the density distribution within the cell and, 
hence, provide an accuracy of second order. Equations~(10) to (12) are convolution forms of double 
summations, which can be evaluated using FFT if the domain is uniformly discretized. Equations~(13) 
to (15) can be integrated analytically and the detailed expressions are summarized in Appendix A.

\subsubsection{Modified particle method}
The 2nd-order scheme described above involves double summations with three types of force kernels, i.e., 
$\mathcal{K}^{x,0}_{i-i',j-j'}$, $\mathcal{K}^{x,x}_{i-i',j-j'}$ and $\mathcal{K}^{x,y}_{i-i',j-j'}$, 
when calculating forces in $x$-direction. The computational cost can be considerably reduced 
if a further approximation is adopted for Equation~(6):
\begin{eqnarray}
F^x_{i,j}&\approx& \sum_{i'=1}^{N_{\rm d}}\sum_{j'=1}^{N_{\rm d}}\frac{x_{i'}-x_i}{[\epsilon^2+(x_{i'}-x_i)^2+(y_{j'}-y_j)^2]^{3/2}}\int\int_{D_{i',j'}}\sigma(\bar{x},\bar{y}){\rm d}\bar{x}{\rm d}\bar{y}, \\
&=& \sum_{i'=1}^{N_{\rm d}}\sum_{j'=1}^{N_{\rm d}}\widetilde{\mathcal{K}}^{x,p}_{i-i',j-j'}M_{i',j'},
\end{eqnarray}
where $\epsilon>0$ denotes the softening length.  Equation (17) is equivalent to placing a particle with 
mass $M_{i',j'}\equiv \int\int_{D_{i',j'}}\sigma(\bar{x},\bar{y}){\rm d}\bar{x}{\rm d}\bar{y}$ at the cell 
center $(x_{i'},y_{j'})$ and the $F^x_{i,j}$ is approximated as a result of a direct summation.
That is, the $N$-body calculation uses $M_{i',j'}=\sigma_{i',j'}\Delta x\Delta y$.  We note that the 
expression of Equation (17) is still a convolution form of double summation with a kernel $\widetilde
{\mathcal{K}}^{x,p}_{i-i',j-j'}\equiv (x_{i'}-x_i)/[\epsilon^2+(x_{i'}-x_i)^2+(y_{j'}-y_j)^2]^{3/2}$. 
A FFT can be applied to reduce the computational cost compared to the use of direct summation. A 
nonzero softening length $\epsilon$ is usually introduced in the denominator of $\widetilde{\mathcal{K}}^
{x,p}_{i-i',j-j'}$ to avoid the singularity when $\vect{x}'=\vect{x}$. This calculation involves only one 
double summation, but at the expense of an order of accuracy. One may expect the calculation of 
Equation (17) is roughly three times faster as compared to that of Equation~(\ref{Eq:Fx_approx}). 

Introducing a softening length is not a desirable feature since forces are distorted, which reduce 
the accuracy of numerical solutions. However, the problem of a singularity does not exist in Equation 
(13) since it is integrable. This suggests a way to avoid the use of a softening length and an 
improvement to the kernel function $\widetilde{\mathcal{K}}^{x,p}_{i-i',j-j'}$ is possible. The improved 
force calculation $F^x_{i,j}$ is proposed as follows:
 \begin{eqnarray}
F^x_{i,j}&\approx& \sigma_{i,j} \mathcal{K}^{x,0}_{0,0}+\delta^x_{i,j} \mathcal{K}^{x,x}_{0,0}+\delta^y_{i,j}\mathcal{K}^{x,y}_{0,0}+\sum_{i'=1}^{N_{\rm d}}\sum_{j'=1}^{N_{\rm d}}\mathcal{K}^{x,p}_{i-i',j-j'}M_{i',j'}, \\
&=&F^{x,corr}_{i,j}+\sum_{i'=1}^{N_{\rm d}}\sum_{j'=1}^{N_{\rm d}}\mathcal{K}^{x,p}_{i-i',j-j'}M_{i',j'},
\end{eqnarray}
where
\begin{eqnarray}
\mathcal{K}^{x,p}_{i-i',j-j'}&=&\begin{cases}
                               \dfrac{x_{i'}-x_i}{[(x_{i'}-x_i)^2+(y_{j'}-y_j)^2]^{3/2}}, & i\neq i'\text{ or } j\neq j' \\
                               0, & \text{otherwise}.
                              \end{cases}, \\
F^{x,corr}_{i,j} &\equiv& \sigma_{i,j} \mathcal{K}^{x,0}_{0,0}+\delta^x_{i,j} \mathcal{K}^{x,x}_{0,0}+\delta^y_{i,j}\mathcal{K}^{x,y}_{0,0}.
\end{eqnarray}
$F^{x,corr}_{i,j}$ is a correction term that takes into account the gravitational force contributed from 
$D_{i,j}$, which is set to be zero in Equation (20), reflecting the idea that a particle does not feel its 
own gravity. On the right hand side of Equation (21), the first term is zero due to the symmetry of the 
cell, the last term is also zero since the integrand in Equation~(15) is an odd function of $\bar{y}$ with 
respect to the cell center. Only the second term that involves the gradient of the surface density 
in the $x$-direction is in general nonzero. Compared to the double summation, the cost of calculating 
$F^{x,corr}_{i,j}$ is very small, but it significantly improves the accuracy of the numerical solution 
as will be shown in Section 5. The particle method that does not suffer from the problem of singularity and 
involves higher order correction for $\vect{x}'=\vect{x}$ is called the modified particle method in this 
paper. 

\subsection{Polar coordinates}
\subsubsection{A direct method of 2nd-order accuracy}
Corresponding to Equation (3) for Cartesian coordinates, the potential function in polar coordinates can 
be expressed as:
\begin{equation}
\Phi(r,\phi,z)=-\int\int\int\mathcal{K}(\bar{r},r,\bar{\phi}-\phi,\bar{z}-z)\rho(\bar{r},\bar{\phi},\bar{z})\bar{r}{\rm d}\bar{r}{\rm d}\bar{\phi}{\rm d}\bar{z},
\end{equation}
where $(r,\phi,z)$ are polar coordinates and $\mathcal{K}(\bar{r},r,\bar{\phi}-\phi,\bar{z}-z)\equiv 1/\sqrt{\bar{r}^2-2r\bar{r}\cos(\bar{\phi}-\phi)+r^2+(\bar{z}-z)^2}$. 

Consider the computational domain described by $\mathcal{R}=\mathcal{R}^{\rm d} \cup \mathcal{R}^{\rm 
s} \cup \hat{\mathcal{R}_j}$, which is the union of three parts $\mathcal{R}^{\rm d}$, $\mathcal{R}^{\rm s}$ 
and $\hat{\mathcal{R}_j}$. Here, $\mathcal{R}^{\rm d}$ represents the destination domain, where the 
resulting gravitational forces include the contributions from the whole computational domain $\mathcal{R}$. 
$\mathcal{R}^{\rm s}$ is the source domain, in which the mass gravitationally influences $R^{\rm d}$. 
$\hat{\mathcal{R}_j}$ contributes the gravitational forces associated with the origin of the calculational domain and its surroundings, 
which is not included in $\mathcal{R}^{\rm d}$ and $\mathcal{R}^{\rm s}$. The discretization of the domains 
$\mathcal{R}^{\rm d}$, $\mathcal{R}^{\rm s}$ and $\hat{\mathcal{R}_j}$ adopted in this work is described 
below. 

$\mathcal{R}^{\rm d}=[M^{\rm d}_{\rm in},M^{\rm d}_{\rm out}]\times[0,2\pi]$ for some number $M^{\rm d}_
{\rm out}>M^{\rm d}_{\rm in}>0$. The radial direction is discretized in logarithmic form and the 
azimuthal direction is evenly subdivided. Namely, for a positive integer $N_{\rm d}$, we define $\Delta 
\phi=2\pi/N_{\rm d}$, $\beta=(M^{\rm d}_{\rm out}/M^{\rm d}_{\rm in})^{1/N_{\rm d}}$, $r_{i+1/2}=
\beta^{i}M^{\rm d}_{\rm in}$, $\phi_{j+1/2}=j\Delta\phi$, $i,j=0,...,N_{\rm d}$, $r_i=(r_{i-1/2}+
r_{i+1/2})/2$ and $\phi_j=(\phi_{j-1/2}+\phi_{j+1/2})/2$, where $i,j=1,...,N_{\rm d}$. The destination 
domain is covered with $N_{\rm d}^2$ cells defined by $\mathcal{R}^{\rm d}_{i,j}=[r_{i-1/2},r_{i+1/2}]
\times[\phi_{j-1/2},\phi_{j+1/2}]$ for $i,j=1,...,N_{\rm d}$. We note that the arrangement of $\mathcal{R}^{\rm d}_{i,j}$ does not cover the region $r<M^{\rm d}_{\rm in}$. Extra cells that cover $r<M^{\rm d}_{\rm in}$ should be included in the region $\mathcal{R}^{\rm s}=[M^{s}_{\rm in},M^{\rm d}_{\rm in}]\times[0,2\pi]$. The region $\mathcal{R}^{\rm s}$ is discretized in the same way used for discretizing $\mathcal{R}^{\rm d}$, i.e., 
using the same $\beta$ and $\Delta \phi$. Without loss of generality, the region $\mathcal{R}^{\rm s}$ is 
discretized with $N_{\rm s}^2$ cells, and with $M^{s}_{\rm in}=\beta^{-N_{\rm s}}M^{\rm d}_{\rm in}$. We 
define $r_{i+1/2}=\beta^{i}M^{\rm d}_{\rm in}$, for $i=-N_{\rm s},...,0$, and $r_i=(r_{i-1/2}+r_{i+1/2})/2$ 
for $i=-N_{\rm s}+1,...,0$. Since the discretization in the azimuthal direction is directly inherited from 
that used for $\mathcal{R}^{\rm d}$, no special care is required. The cells defined by $\mathcal{R}^{\rm s}_
{i,j}=[r_{i-1/2},r_{i+1/2}]\times[\phi_{j-1/2},\phi_{j+1/2}]$ for $i=-N_{\rm s}+1,...,0$ and $j=1,...,N_
{\rm d}$ are used to cover $\mathcal{R}^{\rm s}$.  Finally, cells, $\hat{\mathcal{R}_j}=[0,M^{\rm s}_
{\rm in}]\times[\phi_{j-1/2},\phi_{j+1/2}]$, should be included to take into account the contribution from 
the vicinity around the origin. For simplification of notation, we denote $\mathcal{R}_{-N_{\rm s},j}=\hat
{\mathcal{R}_j}$ and $\mathcal{R}_{i,j}=[r_{i-1/2},r_{i+1/2}]\times[\phi_{j-1/2},\phi_{j+1/2}]$ for the 
ranges of indices $i=-N_{\rm s}+1,...,N_{\rm d}$ and $j=1,...,N_{\rm d}$. 
  
The forces in the $r$-direction ($F^r_{i,j}$) and $\phi$-direction ($F^{\phi}_{i,j}$) are defined at the 
center of cells $(r_i,\phi_j) \in \mathcal{R}^{\rm d}$ and related to Equation (22) through the following 
relations:
\begin{eqnarray}
F^r_{i,j}\equiv-\frac{\partial}{\partial r}\Phi(r_i,\phi_j,0)=\sum^{N_{\rm d}}_{i'=-N_{\rm s}}\sum^{N_{\rm d}}_{j'=1}\int\int_{\mathcal{R}_{i',j'}}\frac{\partial}{\partial r}\mathcal{K}(\bar{r},r_i,\bar{\phi}-\phi_j,0)\sigma({\bar{r},\bar{\phi}})\bar{r}{\rm d}\bar{r}{\rm d}\bar{\phi}, \\
F^{\phi}_{i,j}\equiv-\frac{1}{r_i}\frac{\partial}{\partial \phi}\Phi(r_i,\phi_j,0)=\frac{1}{r_i}\sum^{N_{\rm d}}_{i'=-N_{\rm s}}\sum^{N_{\rm d}}_{j'=1}\int\int_{\mathcal{R}_{i',j'}}\frac{\partial}{\partial \phi}\mathcal{K}(\bar{r},r_i,\bar{\phi}-\phi_j,0)\sigma({\bar{r},\bar{\phi}})\bar{r}{\rm d}\bar{r}{\rm d}\bar{\phi},  
\end{eqnarray}
where the surface density $\sigma(\bar{r},\bar{\phi})$ in $\mathcal{R}_{i',j'}$ is linearly approximated by:
\begin{equation}
\sigma(\bar{r},\bar{\phi})\approx \sigma_{i',j'}+\delta^r_{i',j'}(\bar{r}-r_{i'})+\delta^{\phi}_{i',j'}(\bar{\phi}-\phi_{j'}),
\end{equation}
where $\sigma_{i',j'}$, $\delta^r_{i',j'}\equiv\partial\sigma(r_{i'},\phi_{j'})/\partial r$ and $\delta^{\phi}_{i',j'}\equiv\partial\sigma(r_{i'},\phi_{j'})/\partial \phi$ are constant in the cell $\mathcal{R}_{i',j'}$. With the linear approximation in the surface density, $F^r_{i,j}$ with accuracy of 2nd-order can be approximated by \citep{Yen12}:
\begin{equation}
 F^r_{i,j}\approx F^{r,0}_{i,j}+F^{r,r}_{i,j}+F^{r,\phi}_{i,j},
\end{equation}
where 
\begin{eqnarray}
F^{r,0}_{i,j}&=&\sum^{N_{\rm d}}_{i'=1}\sum^{N_{\rm d}}_{j'=1}\sigma_{i',j'}\mathcal{K}^{r,0}_{i-i',j-j'}+\sum^0_{i'=-N_{\rm s}+1}\sum^{N_{\rm d}}_{j'=1}\sigma_{i',j'}\mathcal{K}^{r,0}_{i-i',j-j'}+\sum^{N_{\rm d}}_{j'=1}\sigma_{-N_{\rm s},j'}\bar{\mathcal{K}}^{r,0}_{i+N_{\rm s},j-j'}, \\
F^{r,r}_{i,j}&=&r_i\left[\sum^{N_{\rm d}}_{i'=1}\sum^{N_{\rm d}}_{j'=1}\delta^r_{i',j'}\mathcal{K}^{r,r}_{i-i',j-j'}+\sum^0_{i'=-N_{\rm s}+1}\sum^{N_{\rm d}}_{j'=1}\delta^r_{i',j'}\mathcal{K}^{r,r}_{i-i',j-j'}+\sum^{N_{\rm d}}_{j'=1}\delta^{r}_{-N_{\rm s},j'}\bar{\mathcal{K}}^{r,r}_{i+N_{\rm s},j-j'}\right],\\
F^{r,\phi}_{i,j}&=&\sum^{N_{\rm d}}_{i'=1}\sum^{N_{\rm d}}_{j'=1}\delta^{\phi}_{i',j'}\mathcal{K}^{r,\phi}_{i-i',j-j'}+\sum^0_{i'=-N_{\rm s}+1}\sum^{N_{\rm d}}_{j'=1}\delta^{\phi}_{i',j'}\mathcal{K}^{r,\phi}_{i-i',j-j'}+\sum^{N_{\rm d}}_{j'=1}\delta^{\phi}_{-N_{\rm s},j'}\bar{\mathcal{K}}^{r,\phi}_{i+N_{\rm s},j-j'},
\end{eqnarray}
and 
\begin{eqnarray}
\mathcal{K}^{r,0}_{i-i',j-j'}&\equiv&-\int\int_{\mathcal{R}_{i',j'}} \frac{\bar{r}(r_i-\bar{r}\cos(\bar{\phi}-\phi_j))}{[\bar{r}^2+r^2_i-2\bar{r}r_i\cos(\bar{\phi}-\phi_j)]^{3/2}}{\rm d}\bar{r}{\rm d}\bar{\phi}, \\
\mathcal{K}^{r,r}_{i-i',j-j'}&\equiv&-\int\int_{\mathcal{R}_{i',j'}} \frac{\bar{r}(r_i-\bar{r}\cos(\bar{\phi}-\phi_j))(\bar{r}-r_{i'})}{r_i[\bar{r}^2+r^2_i-2\bar{r}r_i\cos(\bar{\phi}-\phi_j)]^{3/2}}{\rm d}\bar{r}{\rm d}\bar{\phi}, \\
\mathcal{K}^{r,\phi}_{i-i',j-j'}&\equiv&-\int\int_{\mathcal{R}_{i',j'}} \frac{\bar{r}(r_i-\bar{r}\cos(\bar{\phi}-\phi_j))(\bar{\phi}-\phi_{j'})}{[\bar{r}^2+r^2_i-2\bar{r}r_i\cos(\bar{\phi}-\phi_j)]^{3/2}}{\rm d}\bar{r}{\rm d}\bar{\phi}. 
\end{eqnarray}
On the right hand side of Equations (27) to (29), the first terms are the self-gravitating terms, the 
second terms are the ``gravitational interaction terms" contributed from $\mathcal{R}^{\rm s}$ and the 
last terms are the contributions from the vicinity around the origin $\hat{\mathcal{R}_j}$. Although the 
first two terms can be mathematically combined into a single double summation, in practice, we treat those 
three terms separately. The self-gravitating terms involve integration around the singularity, while 
the interaction terms do not. The third terms, though in a convolution form of a single summation, are not 
compatible with the first two terms and therefore treated separately as well. FFT can be applied to all 
these calculations to reduce the computational cost and keep the complexity at $O(N^2\log_2{N})$.

Some properties regarding Equations (27) to (29) are worth mentioning. First, the order of accuracy relies 
on how accurate the force kernels, i.e., Equations (30) to (32), are evaluated. As pointed out in 
\citet{Yen12}, no closed forms are found for the $\mathcal{K}^{r,0}_{i-i',j-j'}$, $\mathcal{K}^{r,r}_
{i-i',j-j'}$ and $\mathcal{K}^{r,\phi}_{i-i',j-j'}$. The integration involves an elliptic integral that can 
only be evaluated numerically. Moreover, the presence of a singularity function in terms of $\ln(1-
\cos(\phi))$ degrades the order of accuracy to first order. It is therefore desirable to improve the 
accuracy of the force kernels given the fact that Equation (26) involves three integrals, which are 
originally dedicated to reach an accuracy of 2nd-order. These considerations will be addressed in the next 
section. Second, Equations (30) to (32) can be used to evaluate ``gravitational interaction" between 
two grid patches, i.e., the destination patch $\mathcal{R}^{\rm d}$ and the source patch $\mathcal{R}^
{\rm s}$, which are partially or completely separated. As long as the their domain discretization shares 
the same $\beta$ and $\Delta \phi$, FFT can be applied to reduce the computational cost. In the case that 
$\mathcal{R}^{\rm d}$ and $\mathcal{R}^{\rm s}$ are completely separated, we do not need to worry about 
the singularity associated with $\ln(1-\cos(\phi))$. We note that a constant spatial shift between two 
patches in $r$ and $\phi$ is allowed, since the spatial shift only contributes constant phase shifts in 
the Fourier domain. Third, in the case that $\mathcal{R}_{i,j}=\mathcal{R}_{i',j'}$, where the 
singularity occurs, the values of  $\mathcal{K}^{r,0}_{0,0}$, $\mathcal{K}^{r,r}_{0,0}$ and 
$\mathcal{K}^{r,\phi}_ {0,0}$ are invariant. For example, casting Equation (31) in the form
\begin{equation}
\mathcal{K}^{r,r}_{0,0}=-\int^{\Delta \phi/2}_{-\Delta \phi/2}\int^{2\beta/(\beta+1)}_{2/(\beta+1)}\frac{\eta(1-\eta\cos(\xi))(\eta-1)}{[\eta^2+1-2\eta\cos(\xi)]^{3/2}}{\rm d}{\eta}{\rm d}\xi, 
\end{equation}
where $\eta\equiv\bar{r}/r_i$ and $\xi\equiv\bar{\phi}-\phi_j$, $\mathcal{K}^{r,r}_{0,0}$ is a constant 
for all $i'=i$ and $j'=j$ since $\Delta \phi$ and $\beta$ are constants. This is a useful property given 
that the cell sizes are not uniform in polar coordinates. This indicates that one can place the effort 
on evaluating the elliptical integral for one specific cell which contains a singularity and apply the 
result to all other cells that also contain a singularity. 

\subsubsection{Modified Particle Method}
The modified particle-based method can also be applied to polar coordinates. Similar to Equation~(16), 
we can approximate Equation~(23) to further reduce the computational cost at the expense of the order 
of convergence. Corresponding to Equation (18) for Cartesian coordinates, the approximation in polar 
coordinates is written as:
\begin{eqnarray}
F^{r}_{i,j}&\approx& \sigma_{i,j}\mathcal{K}^{r,0}_{0,0}+r_i\delta^{r}_{i,j}\mathcal{K}^{r,r}_{0,0}+\delta^{\phi}_{i,j}\mathcal{K}^{r,\phi}
_{0,0}+\sum^{N_{\rm d}}_{i'=-N_{\rm s}}\sum^{N_{\rm d}}_{j'=1}\mathcal{K}^{r,p}_{i-i',j-j'}M_{i',j'}, \\
&=& F^{r,corr}_{i,j}+\sum^{N_{\rm d}}_{i'=-N_{\rm s}}\sum^{N_{\rm d}}_{j'=1}\mathcal{K}^{r,p}_{i-i',j-j'}M_{i',j'}, 
\end{eqnarray}
where
\begin{eqnarray}
\mathcal{K}^{r,p}_{i-i',j-j'}&=&\begin{cases}
                               -\dfrac{(r_i-r_{i'}\cos(\phi_{j'}-\phi_j))}{[r_{i'}^2+r_i^2-2r_{i'}r_i\cos(\phi_{j'}-\phi_j)]^{3/2}}, & i\neq i'\text{ or } j\neq j' \\
                               0, & \text{otherwise}.
                              \end{cases}, \\
F^{r,corr}_{i,j} &\equiv& \sigma_{i,j}\mathcal{K}^{r,0}_{0,0}+r_i\delta^{r}_{i,j}
\mathcal{K}^{r,r}_{0,0}+\delta^{\phi}_{i,j}\mathcal{K}^{r,\phi} _{0,0}.
\end{eqnarray}
Here, $F^{r,corr}_{i,j}$ denotes the gravitational force contributed from $\mathcal{R}_{i,j}$, which is 
set to zero in Equation~(36).  It can be also shown that $\mathcal{K}^{r,\phi}_{0,0}=0$ since the integrand 
in Equation~(32) is an odd function of $\bar{\phi}$ with respect to the cell center. The first two terms 
on the right hand side of Equation~(37) are in general nonzero. Since the shape of cells in polar 
coordinates is not symmetric with respect to the center of cells, involving the correction term 
$\sigma_{i,j}\mathcal{K}^{r,0}_{0,0}$ is particularly relevant in polar coordinates. The cost of computing 
$F^{r,corr}_{i,j}$ is small compared to the calculation of a double summation. 

\section{Singular integration method (SIM)}
The mathematical formulas developed for 2nd-order convergence have been shown in Section 3.2. The lack of 
closed forms for Equations~(30) to (32) dictates that the order of convergence relies on 
the numerical methods used for the integrations. \citet{Yen12} evaluate the force kernels using the 
trapezoidal rule with one trapezoid as follows 
\begin{eqnarray}
\mathcal{K}^{r,0}_{i-i',j-j'} &\approx & -\mathcal{H}^r_1\left(\frac{\bar{r}}{r},\bar{\phi}-\phi_j\right) \left|^{r_{i'+1/2}}_{r_{i'-1/2}}\right.\left.\right]^{\phi_{j'+1/2}}_{\phi_{j'-1/2}}, \\
\mathcal{K}^{r,r}_{i-i',j-j'} &\approx & -\mathcal{H}^r_2\left(\frac{\bar{r}}{r}, \bar{\phi}-\phi_j \right)\left|^{r_{i'+1/2}}_{r_{i'-1/2}}\right.\left.\right]^{\phi_{j'+1/2}}_{\phi_{j'-1/2}}-\frac{r_{i'}}{r_i}\mathcal{K}^{r,0}_{i-i',j-j'}, \\
\mathcal{K}^{r,\phi}_{i-i',j-j'} &\approx & -(\bar{\phi}-\phi_{j'})\mathcal{H}^r_1\left(\frac{\bar{r}}{r},\bar{\phi}-\phi_j\right) \left|^{r_{i'+1/2}}_{r_{i'-1/2}}\right.\left.\right]^{\phi_{j'+1/2}}_{\phi_{j'-1/2}}, 
\end{eqnarray}
where the notation $f(\cdot)]^b_a\equiv(f(b)+f(a))(b-a)/2$, and the exact expressions of 
$\mathcal{H}^r_1$ and $\mathcal{H}^r_2$ are given in Appendix A. It is natural to utilize more than 
one trapezoid to improve the accuracy. Specifically, 
\begin{eqnarray}
\mathcal{K}^{r,0}_{i-i',j-j'} &\approx & \sum^{N_{\rm tpz}}_{m=1}-\mathcal{H}^r_1\left(\frac{\bar{r}}{r},\bar{\phi}-\phi_j\right) \left|^{r_{i'+1/2}}_{r_{i'-1/2}}\right.\left.\right]^{\phi'_{m+1}}_{\phi'_{m}}, \\
\mathcal{K}^{r,r}_{i-i',j-j'} &\approx & \sum^{N_{\rm tpz}}_{m=1} -\mathcal{H}^r_2\left(\frac{\bar{r}}{r}, \bar{\phi}-\phi_j \right)\left|^{r_{i'+1/2}}_{r_{i'-1/2}}\right.\left.\right]^{\phi'_{m+1}}_{\phi'_{m}}-\frac{r_{i'}}{r_i}\mathcal{K}^{r,0}_{i-i',j-j'}, \\
\mathcal{K}^{r,\phi}_{i-i',j-j'} &\approx & \sum^{N_{\rm tpz}}_{m=1}-(\bar{\phi}-\phi_{j'})\mathcal{H}^r_1\left(\frac{\bar{r}}{r},\bar{\phi}-\phi_j\right) \left|^{r_{i'+1/2}}_{r_{i'-1/2}}\right.\left.\right]^{\phi'_{m+1}}_{\phi'_{m}}, 
\end{eqnarray}
where $\phi'_m=\phi_{j'-1/2}+(m-1)\Delta \theta$, $\Delta \theta=(\phi_{j'+1/2}-\phi_{j'-1/2})/N_{\rm tpz}$ and $N_{\rm tpz}>1$ denotes the number of trapezoid used for the evaluation.  As will be shown in Section 5, this consideration significantly improves the accuracy of numerical solutions in the self-gravitating case. We emphasize that evaluating Equations~(41) to (43) using $N_{\rm tpz}>1$ does not increase the computational complexity of the method, since those kernels are calculated only once at the beginning of simulations. 

Another issue associated with Equations~(30) to (32) is that when $i=i'$ and $j=j'$, a singularity in the 
form of $\ln(1-\cos(\phi))$ is involved in those integrals. In this situation, forces evaluated using 
Equations~(38) to (40) can have incorrect signs, 
which not only degrades the order of convergence to first order but also deteriorates the accuracy of 
numerical solutions. Thus, special care is required for the evaluation of $\mathcal{K}^{r,0}_{0,0}$, 
$\mathcal{K}^{r,r}_{0,0}$, and $\mathcal{K}^{r,\phi}_{0,0}$. Fortunately, Equation~(33) suggests that 
special care need only to be taken once for one specific cell and the result can be applied to other cells. 

\begin{figure}
\epsscale{1}
\plotone{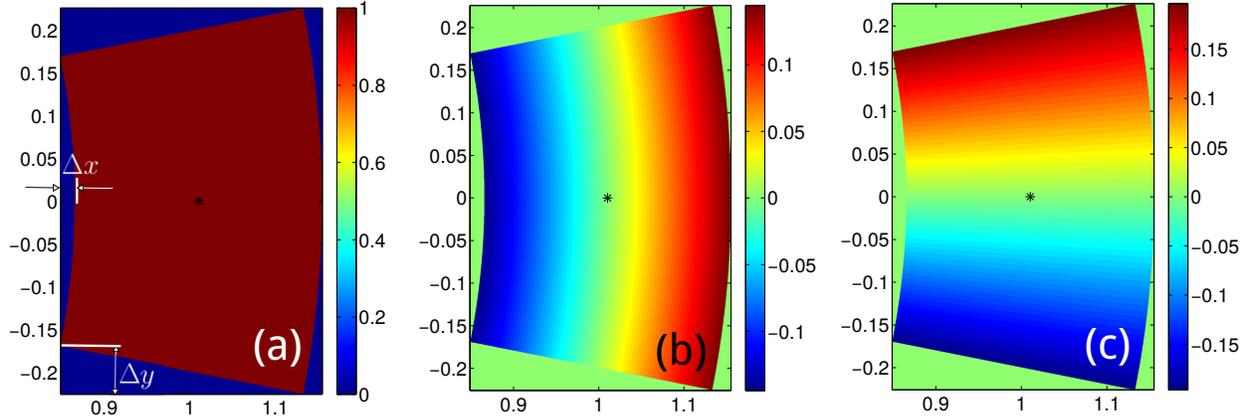}
\caption{Evaluation of $\mathcal{K}^{r,0}_{0,0}$, $\mathcal{K}^{r,r}_{0,0}$ and $\mathcal{K}^{\phi,\phi}_{0,0}$ at the cell center, $(r_m,\phi_m)$, as denoted by the asterisk symbols. The fan-shaped cell is covered with Cartesian cells. The forces are evaluated using the 2nd-order scheme described in Section 3.1 to avoid the singularity appearing in polar coordinates. (a) Surface density with $\sigma=1$ is used to evaluate $\mathcal{K}^{r,0}_{0,0}$. (b) Surface density with unit slope in radial direction is used to evaluate $r_m \mathcal{K}^{r,r}_{0,0}$. (c) Surface density with unit slope in azimuthal direction is used to evaluate $\mathcal{K}^{\phi,\phi}_{0,0}$. The fan-shaped cell, which is characterized by $\Delta x$ and $\Delta y$, should be spatially resolved using roughly 10 Cartesian cells to have a reasonable speed of convergence.\label{fig:FanKernel}}
\end{figure}

As shown in Figure~\ref{fig:FanKernel}(a), we cover a specific fan-shaped cell using Cartesian cells. 
The fan-shaped cell can be characterized by:
\begin{eqnarray}
\Delta x&=&\left(r_m-\frac{\Delta r}{2}\right)\left[1-\cos\left(\frac{\Delta \phi}{2}\right)\right], \\
\Delta y&=&\Delta r \sin\left(\frac{\Delta \phi}{2}\right),
\end{eqnarray}
where $(r_m,\phi_m=0)$ denotes the cell center and $(\Delta r, \Delta \phi)$ defines the size of the fan-shaped cell. Since $(\Delta r, \Delta \phi)$ decreases with increasing $N_{\rm d}$, the number of Cartesian cells used to cover the fan-shaped area should increase accordingly to well resolve the curved fan-shape. A good rule of thumb is to cover $\Delta x$ and $\Delta y$ with roughly 10 Cartesian cells in $x$ and $y$ direction, respectively. As a result, the corresponding number of Cartesian cells used to cover the fan-shaped cell in this work is $10N_{\rm d}$ in the $x$ direction and $5N_{\rm d}$ in the $y$ direction. The surface density of those Cartesian cells that lie outside the fan-shaped area is set to zero. 

We evaluate the forces at the cell center, $(r_m,\phi_m)$, associated with (a) uniform surface density $\sigma(\bar{x},
\bar{y})=1$, (b) $\sigma(\bar{x},\bar{y})=\bar{r}-r_m$ and (c) $\sigma(\bar{x},\bar{y})=\bar{\phi}-\phi_m$. 
Since we are only interested in the self-gravitating forces at a specific point, $(r_m,\phi_m)$, evaluation using Equation~(9) that involves only three double summations would suffice. 
The force calculated for Figure~\ref{fig:FanKernel}(a) corresponds to 
$\mathcal{K}^{r,0}_{0,0}$ due to unit surface density, Figure~\ref{fig:FanKernel}(b) corresponds to 
$r_m\mathcal{K}^{r,r}_{0,0}$ due to the unit radial slope in surface density, and Figure~\ref{fig:FanKernel}
(c) corresponds to $\mathcal{K}^{r,\phi}_{0,0}=0$ due to the unit azimuthal slope in surface density.  
Figure~\ref{fig:FanKernel}(c) is more relevant to $\mathcal{K}^{\phi,\phi}_{0,0}$ as shown in Appendix A. 
We note that when applying the force, $F^{r,r}_{r_m,\phi_m}$, associated with Figure~\ref{fig:FanKernel}(b) 
to other cells, a factor $r_i/r_m$ is required, i.e., 
\begin{equation}
F^{r,r}_{i,j}=\frac{r_i}{r_m}F^{r,r}_{r_m,\phi_m}=r_i\mathcal{K}^{r,r}_{0,0},  
\end{equation}
due to the factor $r_i$ appearing in front of the square bracket in Equation~(28). Hereafter, we call 
the algorithm described in this section as the singularity integration method or SIM in short. 
 
\section{Results}
In this section, we verify the accuracy and the order of convergence proposed in this work by comparing the 
numerical solutions with examples that have an analytic solutions. For Cartesian coordinates, we focus on 
the accuracy improvement for the modified particle method, while we show improvements in both the 
accuracy and the order of convergence for polar coordinates. 

We investigate the numerical error in the destination domain $\mathcal{R}^{\rm d}$ and $\mathcal{D}$ using 
the following definitions of error:
\begin{eqnarray}
  L^1 &=& \frac{1}{N_{\rm d}^2}\sum_{i=1}^{N_{\rm d}}\sum_{j=1}^{N_{\rm d}} \left| F^{{\rm num}}_{i,j}-F^{{\rm ana}}_{i,j}\right|, \\
  L^2 &=& \left(\frac{1}{N_{\rm d}^2}\sum_{i=1}^{N_{\rm d}}\sum_{j=1}^{N_{\rm d}} \left| F^{{\rm num}}_{i,j}-F^{{\rm ana}}_{i,j}\right|^2\right)^{1/2}, \\
  L^{\infty}&=& {\rm max}(\left| F^{{\rm num}}_{i,j}-F^{{\rm ana}}_{i,j}\right|) \quad\quad \text{for~}i,j\in \mathcal{R}^{\rm d}~\text{or~}\mathcal{D},
\end{eqnarray}
where $L^1$, $L^2$, $L^{\infty}$ are the one norm, two norm and maximum norm of error, and 
$F^{{\rm num}}_{i,j}$, $F^{{\rm ana}}_{i,j}$ are numerical and exact forces at locations indexed by 
$(i,j)$, respectively. When using $L^1$ and $L^2$, we evaluate the total variation and energy in a global 
sense, while using $L^{\infty}$ we focus on the convergence of maximum error in a pointwise sense. 
\subsection{Examples with analytic solutions}
Direct comparisons between numerical with analytical solutions are desirable to demonstrate  
the effectiveness of a numerical method. In this work, we are concerned with the accuracy and the order of convergence of the proposed algorithms. For these purposes, the selected disk models need to fulfill the following criteria. First, the disk is supposed to be infinitesimally thin and has closed-form solutions for the self-gravitational forces and the potential. Second, the size of the disk should be finite to be fully enclosed by the finite calculation domain. Third, the mathematical form of the density distribution should be sufficiently smooth, i.e., higher order derivatives behave well in the calculation domain, for analyzing the order of accuracy. Disk models that do not fulfill these criteria are not useful in understanding the properties of the proposed algorithms.

 Only few infinitesimally thin disks  are found to have corresponding closed-form solutions of self-gravitational  forces, e.g., the Mestel disks \citep{Mes63}, the exponential disks and the generalized Maclaurin disks \citep{Sch09}.  Among them, only the density-potential pairs discussed by \citet{Sch09} satisfy the first two criteria above mentioned. \citet{Sch09} found closed-form solutions in cylindrical coordinates for the first three members $n=0,1,2$ of the family of finite disks with surface density, $\sigma_{D_n}$, described by:
\begin{eqnarray}
\sigma_{D_n}(r;\alpha)=\begin{cases}
                               \sigma_0\left(1-\dfrac{r^2}{\alpha^2}\right)^{n-1/2}, & \text{~for~}r<\alpha \\
                               0, & \text{~for~}r\geq \alpha, 
                              \end{cases},
\end{eqnarray} 
 where $r=\sqrt{x^2+y^2}$ and $\alpha$ is a given constant describing the size of disks. It can be shown that even the $\sigma_{D_2}$ disk, which is smoothest among the three, has a singularity in the second derivative of Equation~(50) at $r=\alpha$. In other words, $\sigma_{D_2}$ disk is not sufficiently smooth along the disk edge and will degrade the order of convergence in terms of the maximum norm error. To circumvent this issue,  we generalize the closed-form solutions for arbitrary positive integer $n>0$ as the following. 
 
In general, for a given integer $n \ge 0$, the surface density of $\sigma_{D_{n+1}}$ can be associated with $\sigma_{D_n}$ though the following recursive relation:
\begin{equation}
   \sigma_{D_{n+1}}(r;\alpha)=\frac{2n+1}{\alpha^{2n+1}}\int_0^{\alpha}\hat{\alpha}^{2n}\sigma_{D_n}(r; \hat{\alpha}){\rm d}{\hat{\alpha}},
\end{equation}
where $\hat{\alpha}$ serves as a dummy variable for the integration. Due to the linearity of the Poisson equation,  the corresponding radial force at the mid-plane has a similar recursive relation:
\begin{equation}
   F^{r,{\rm ana}}_{D_{n+1}}(r;\alpha)=\frac{2n+1}{\alpha^{2n+1}}\int_0^{\alpha}\hat{\alpha}^{2n}F^{r,{\rm ana}}_{D_n}(r; \hat{\alpha}){\rm d}{\hat{\alpha}},
\end{equation}

Without loss of generality for the discussion in this work, we assume $\sigma_0=1$ and $G=1$. The closed-form of the radial force $F^{r,{\rm ana}}_{D_{n+1}}$ has the following general form:
\begin{eqnarray}
F^{r,{\rm ana}}_{n}(r;\alpha)=\begin{cases}
                               \dfrac{\pi^2}{2}\left(\dfrac{r}{\alpha}\right)^{2n-1}\left[\displaystyle \sum^{n-1}_{k=0}b^{n}_{2k}T_{2k}\left(\dfrac{\alpha}{r}\right)\right], & \text{~for~}r\leq\alpha \\
                               \pi\left(\dfrac{r}{\alpha}\right)^{2n-1}\left[\displaystyle\sum^{n-1}_{k=0}b^{n}_{2k}T_{2k}\left(\dfrac{\alpha}{r}\right)\sin^{-1}\left(\dfrac{\alpha}{r}\right) \right. \\
                               \left. + \displaystyle\sum^{n-1}_{k=0}a^n_{2k+1}T_{2k+1}\left(\dfrac{\alpha}{r}\right)\sqrt{1-(\alpha/r)^2}\right],  & \text{~for~}r\geq \alpha 
                              \end{cases}
\end{eqnarray}
where $T_{2k}$ is the Chebyshev polynomial of the first kind of order $2k$ and ($a^n_{2k+1}$, $b^n_{2k}$) are the coefficients associated with the odd and the even order of Chebyshev polynomials, respectively. The coefficients $(a^{n+1},b^{n+1})$ have the following recursive relation with $(a^{n}, b^{n})$:
\begin{eqnarray}
\begin{cases}
a^{n+1}_{2j+1}=(2n+1)\mathcal{F}, & j=1,...,n \\
a^{n+1}_1=(2n+1)\left(\mathcal{F}+\dfrac{b^n_0}{8}+\dfrac{a^n_1}{8}\right), & j=0 \\
\mathcal{F}= -\dfrac{1}{4}\displaystyle\sum^{n}_{k=j} \dfrac{a^{n}_{2k+1}}{2k+2}+\dfrac{1}{4}\displaystyle\sum^{n}_{k=j}\dfrac{a^n_{2k-1}}{2k+2}-\dfrac{1}{4}\displaystyle\sum^{n-2}_{k=j}\dfrac{a^2_{2k+3}}{2k+2} \\
+\dfrac{1}{4}\displaystyle\sum^{n-3}_{k=j}\dfrac{a^n_{2k+5}}{2k+2}+\dfrac{1}{2}\displaystyle\sum^{n-1}_{k=j}\dfrac{b^n_{2k}}{(2k+2)^2}-\dfrac{1}{2}\displaystyle\sum^{n-3}_{k=j}\dfrac{b^n_{2k+4}}{(2k+2)^2}, & j=0,...,n
\end{cases}
\end{eqnarray}
and 
\begin{eqnarray}
\begin{cases}
b^{n+1}_{2j}=(2n+1)\mathcal{G}, & j=2,...,n \\
b^{n+1}_{2}=(2n+1)\left(\mathcal{G}+\dfrac{b^n_0}{8}\right), & j=1 \\
b^{n+1}_0=(2n+1)\left(\dfrac{a^n_1}{8}-\dfrac{a^n_3}{8}\right), & j=0 \\
\mathcal{G}=\dfrac{1}{8}\left(\dfrac{b^n_{2j-2}}{j}-\dfrac{b^n_{2j+2}}{j} \right), & j=1,...,n
\end{cases}
\end{eqnarray}
with $a^1_1=1$ and $b^1_0=-1$. The derivation of Equations~(54), (55) are detailed in Appendix B. By using a $\sigma_{D_n}$ disk, the surface density has  smoothed $n-1$ order of derivative at the edge of the disk. In the following, we adopt $n=2$ and 5 in illustrating the issue associated with the smoothness of the surface density, i.e., the third criterion, and justify that the SIM is of nearly 2nd-order accuracy. Since the Poisson equation is linear and those examples considered in this paper involves all Fourier modes in both the radial and azimuthal directions, the conclusions drawn from this work are general and applicable to any other smooth density distribution. 

\subsection{Results of Cartesian coordinates}
\begin{figure}
\epsscale{1.1}
\plotone{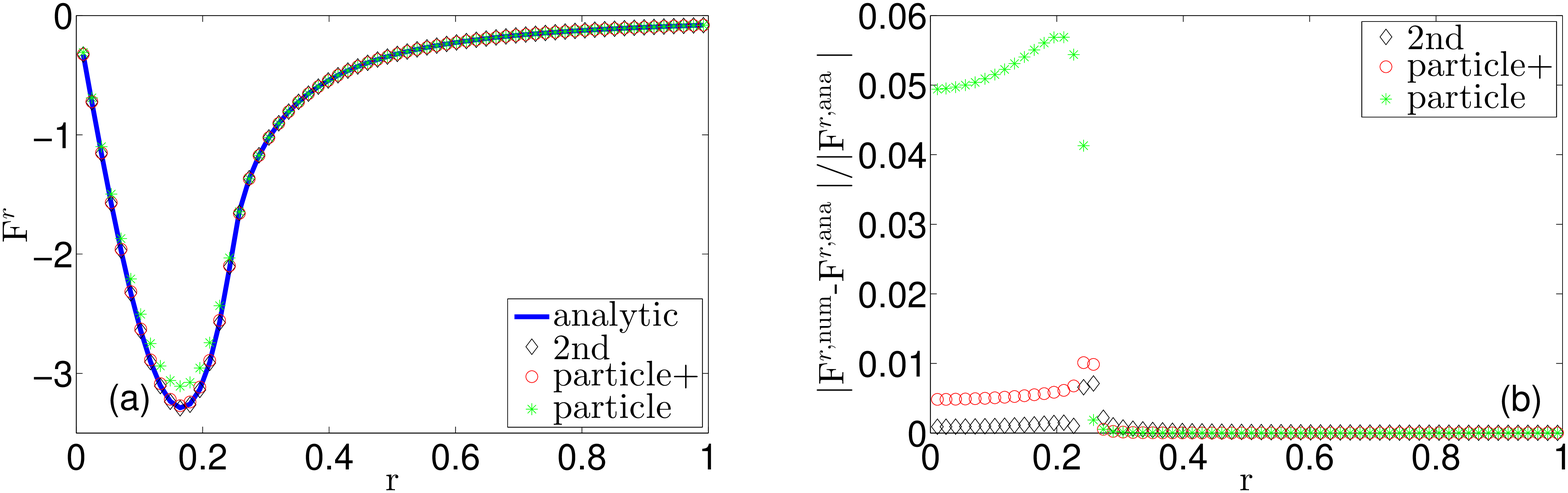}
\caption{Comparisons of radial forces for the $\sigma_{D_2}$ model with $\alpha=0.25$ and $N=128$ in Cartesian coordinates. The solid line is the analytic solution, diamond symbol is the solution from the 2nd-order scheme, empty circle (particle+) is obtained using particle-based method with $\mathcal{K}^{x,x}_{0,0}$ and $\mathcal{K}^{y,y}_{0,0}$ correction and asterisk (particle) is the particle-based method without density slope correction. (a) Radial forces as a function of radius. (b) Relative error as a function of radius. \label{fig:Fr_err}}
\end{figure}

The comparisons of the calculated radial force in Cartesian coordinates for different methods are 
shown in Figure~\ref{fig:Fr_err} for $\sigma_{D_2}$ with $\alpha=0.25$ using $N=128$. In Figure~\ref{fig:Fr_err}(a), the 2nd-order method and the modified particle method (denoted as particle+) have better 
numerical accuracy compared to that of particle method without correction (denoted as particle). The 
absolute value of the radial force is significantly underestimated in the last case, which neglects the 
density gradient in one cell. Since the gradient is negative, i.e., density is higher toward the center of 
the disk, one may expect a radially inward force contributed from inside a cell. The improvement is best 
shown in Figure~\ref{fig:Fr_err}(b) using relative error defined by $|F^{r,{\rm num}}-F^{r,{\rm ana}}|/
|F^{r,{\rm ana}}|$. Compared to the particle method without correction, the absolute error is reduced by an 
order of magnitude if the density slope within the cell is taken into account, and an additional factor of 
five improvement is found in the 2nd-order method. This is significant due to the inverse square 
law of the gravitational force. 

\begin{figure}
\epsscale{1}
\plotone{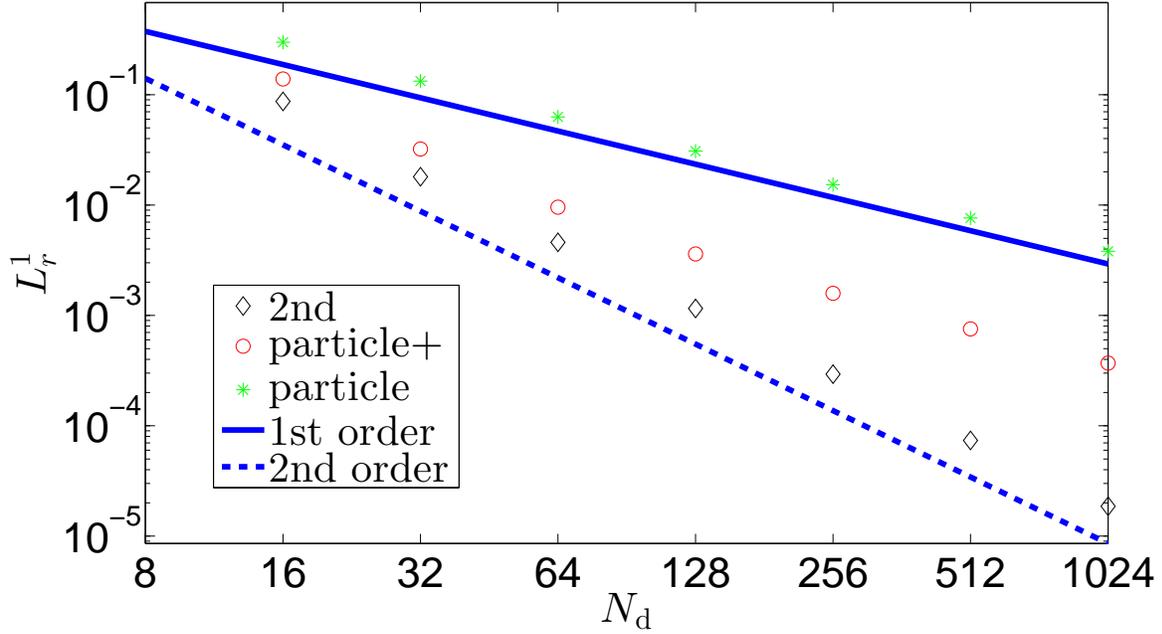}
\caption{The one norm error, $L^1_{r}$, as a function of cell number $N$ for the $\sigma_{D_2}$ model with $\alpha=0.25$ in Cartesian coordinates. The diamond symbol is the error from the 2nd-order scheme, empty circle (particle+) is obtained using particle-based method with density slope correction and asterisk (particle) is the particle-based method without density slope correction. The solid and the dashed lines indicate the slopes of 1st-order and 2nd-order convergence, respectively. The order of convergence fitted for the last four data points are 2.0, 1.1, 1.0 for the 2nd-order, the particle+ and the particle methods, respectively.  \label{fig:order_acc_alpha025_Cartesian}}
\end{figure}

Figure~\ref{fig:order_acc_alpha025_Cartesian} shows the one norm error of radial forces ($L^1_r$) as a 
function of cell number $N_{\rm d}$. The decrement in $L^1_r$ with increasing $N_{\rm d}$ indicates the 
convergence of all methods. The slope corresponds to the order of convergence as indicated by the solid 
line for the 1st-order and the dashed line for the 2nd-order. This figure shows that, in general, the 
2nd-order scheme is the most accurate method compared to others and indeed has numerical convergence 
of nearly 2nd order. Although both the particle-based methods have numerical convergence of nearly 
first order, the numerical accuracy of modified particle method is better than that of the particle method 
by one order of magnitude. 

\subsection{Results of polar coordinates}
We have three cases for cylindrical coordinates. In the first case, we demonstrate the order of convergence 
of the methods in the absence of a singularity. To do so, a $\sigma_{D_2}$ disk with $\alpha=0.006$ 
centered at the origin is employed in the source domain $\mathcal{R}^{\rm s}=[M^{\rm s}_{\rm in}, 
10^{-2}]\times[0,2\pi]$ and $\hat{\mathcal{R}}=[0,M^{\rm s}_{\rm in}]\times[0,2\pi]$ that encloses the 
origin, keeping the destination domain $\mathcal{R}^{\rm d}=[10^{-2},1]\times[0,2\pi]$ devoid of mass. 
We adopt $M^{\rm s}_{\rm in}=10^{-4}$ when $N_{\rm d}=8$ and the value of $M^{\rm s}_{\rm in}$ consecutively 
shrinks to roughly half the size whenever $N_{\rm d}$ is doubled. 

\begin{figure}
\epsscale{1}
\plotone{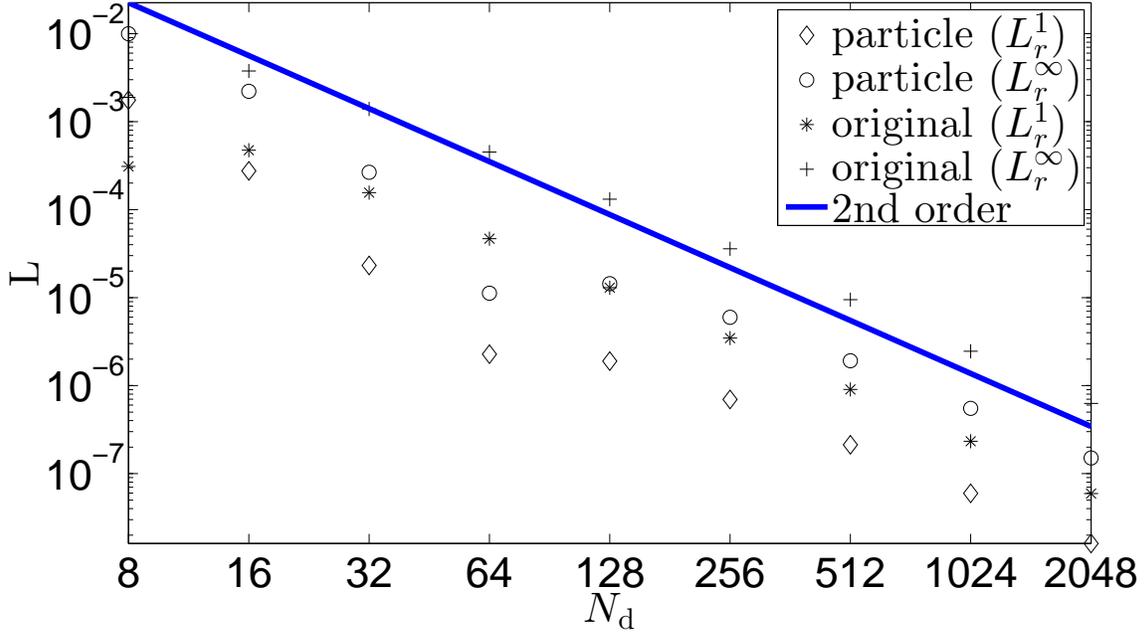}
\caption{The one norm and maximum norm errors as a function of cell number $N$ for the $\sigma_{D_2}$ model with $\alpha=0.006$ in cylindrical coordinates. The diamonds and open circles are obtained from the particle-based method, while the asterisks and plus signs are from the original method proposed in \citet{Yen12}, i.e., $N_{\rm tpz}=1$. The solid line indicates the slope of 2nd-order convergence. The order of convergence fitted for the last four data points are 1.8, 1.8, 2.0, 2.0 for the diamond, the open circle, the asterisk and the plus sign data, respectively. \label{fig:particle_order}}
\end{figure}

Figure~\ref{fig:particle_order} shows the one norm and the maximum norm errors of the radial forces as a 
function of $N_{\rm d}$. The convergence of both the particle method and the original method proposed 
by \citet{Yen12} are of nearly 2nd-order. The 2nd-order convergence of the particle method is general 
if the destination domain $\mathcal{R}^{\rm d}$ is devoid of mass to avoid the singularity involved in the 
integrals Equations~(30) to (32). This can be understood as described below.

Consider the location $(r_i,\phi_j) \in \mathcal{R}^{\rm d}$ where one feels the gravitational force 
in the radial direction from $\mathcal{R}^{\rm s}_{i',j'}$:

\begin{eqnarray}
F^r_{i,j; i', j'}=-\int\int_{\mathcal{R}_{i',j'}}\sigma(r',\phi')\frac{r'[r_i-r'\cos(\phi'-\phi_j)]}{[r'^2+r_i^2-2r'r_i\cos(\phi'-\phi_j)]^{3/2}}{\rm d}r'{\rm d}{\phi'}
\end{eqnarray}
substituting the following relations into Equation~(56):
\begin{eqnarray}
r'&=&r_{i'}+\bar{r}, \\
\phi' &=& \phi_{j'}+\bar{\phi}, \\
\sigma_(r',\phi')&\approx&\sigma_{i',j'}+\delta^r_{i',j'}{(r'-r_{i'})}+\delta^{\phi}_{i',j'}(\phi-\phi_{j'}) \nonumber \\
                 &=&\sigma_{i',j'}+\delta^r_{i',j'}\bar{r}+\delta^{\phi}_{i',j'}\bar{\phi}, \\
\cos(\phi'-\phi_j)&=& \cos(\bar{\phi}+\phi_{j'}-\phi_j)=\cos(\phi_{j'}-\phi_j)\left[1-\frac{\bar{\phi}^2}{2}+O(\bar{\phi}^4)\right] \nonumber \\
                  &-& \sin(\phi_{j'}-\phi_j)\left[\bar{\phi}-\frac{\bar{\phi}^3}{6}+O(\bar{\phi}^5)\right]
\end{eqnarray}
It can be shown that:
\begin{eqnarray}
F^{r}_{i,j;i',j'}&=&\frac{-r_{i'}[r_i-r_{i'}\cos(\phi_{j'}-\phi_j)]}{[r_{i'}^2+r_i^2-2r_{i'}r_i\cos(\phi_{j'}-\phi_j)]^{3/2}}\int\limits^{\Delta\phi_{j'}/2}_{-\Delta\phi_{j'}/2}\int\limits^{\Delta r_{i'}/2}_{-\Delta r_{i'}/2}(\sigma_{i',j'}+\delta^{r}_{i',j'}\bar{r}+\delta^{\phi}_{i',j'}\bar{\phi})(1+\gamma_1\bar{r} \nonumber \\
&+&\gamma_2\bar{\phi}+\gamma_3\bar{r}^2+\gamma_4\bar{\phi}^2+\gamma_5\bar{r}\bar{\phi}+O(\bar{r}\bar{\phi}^2)){\rm d}\bar{r}{\rm d}\bar{\phi} \\
&=&\frac{-r_{i'}[r_i-r_{i'}\cos(\phi_{j'}-\phi_j)]}{[r_{i'}^2+r_i^2-2r_{i'}r_i\cos(\phi_{j'}-\phi_j)]^{3/2}} \left[\sigma_{i',j'}\Delta r_{i'}\Delta\phi_{j'}(1+O((\Delta r_{i'})^2+(\Delta \phi_{j'})^2)) \nonumber \right. \\
&&\left. +\delta^{r}_{i',j'}\Delta r_{i'}\Delta\phi_{j'}(O((\Delta r_{i'})^2+(\Delta \phi_{j'})^2)))+\delta^{\phi}_{i',j'}\Delta r_{i'}\Delta\phi_{j'}(O((\Delta r_{i'})^2+(\Delta \phi_{j'})^2)))\right] \nonumber \\
&=&\frac{-[r_i-r_{i'}\cos(\phi_{j'}-\phi_j)]}{[r_{i'}^2+r_i^2-2r_{i'}r_i\cos(\phi_{j'}-\phi_j)]^{3/2}}\bar{M}_{i',j'}+O((\Delta r_{i'})^2+(\Delta \phi_{j'})^2)), 
\end{eqnarray}
where $\bar{M}_{i',j'}\equiv \sigma_{i',j'}r_{i'}\Delta r_{i'}\Delta\phi_{j'}$. Equation~(62) indicates that 
the accuracy of the particle-based method is in general of 2nd-order in the absence of a 
singularity. Equation~(61) involves the Taylor expansion of the denominator in Equation~(56). When 
approaching Equation~(61) from Equation~(56), we have to assume that the distance $|\vect{x}-\vect{x}'| 
\gg \Delta r_{i'}$ in order to have a reasonable speed of convergence. This assumption breaks down in 
the self-gravitating case that involves an integration around a singularity that reduces the order 
of accuracy. Figure~\ref{fig:particle_order} also shows that the particle method seems to be more accurate 
than the original method proposed in \citet{Yen12}. The difference comes from the use of central difference 
at the disk edge. The disk mass does not vanish to zero at $r=\alpha$ in the latter case. In spite of this, the 
error is reduced commensurate with that of 2nd-order. 

\begin{figure}
\epsscale{1}
\plotone{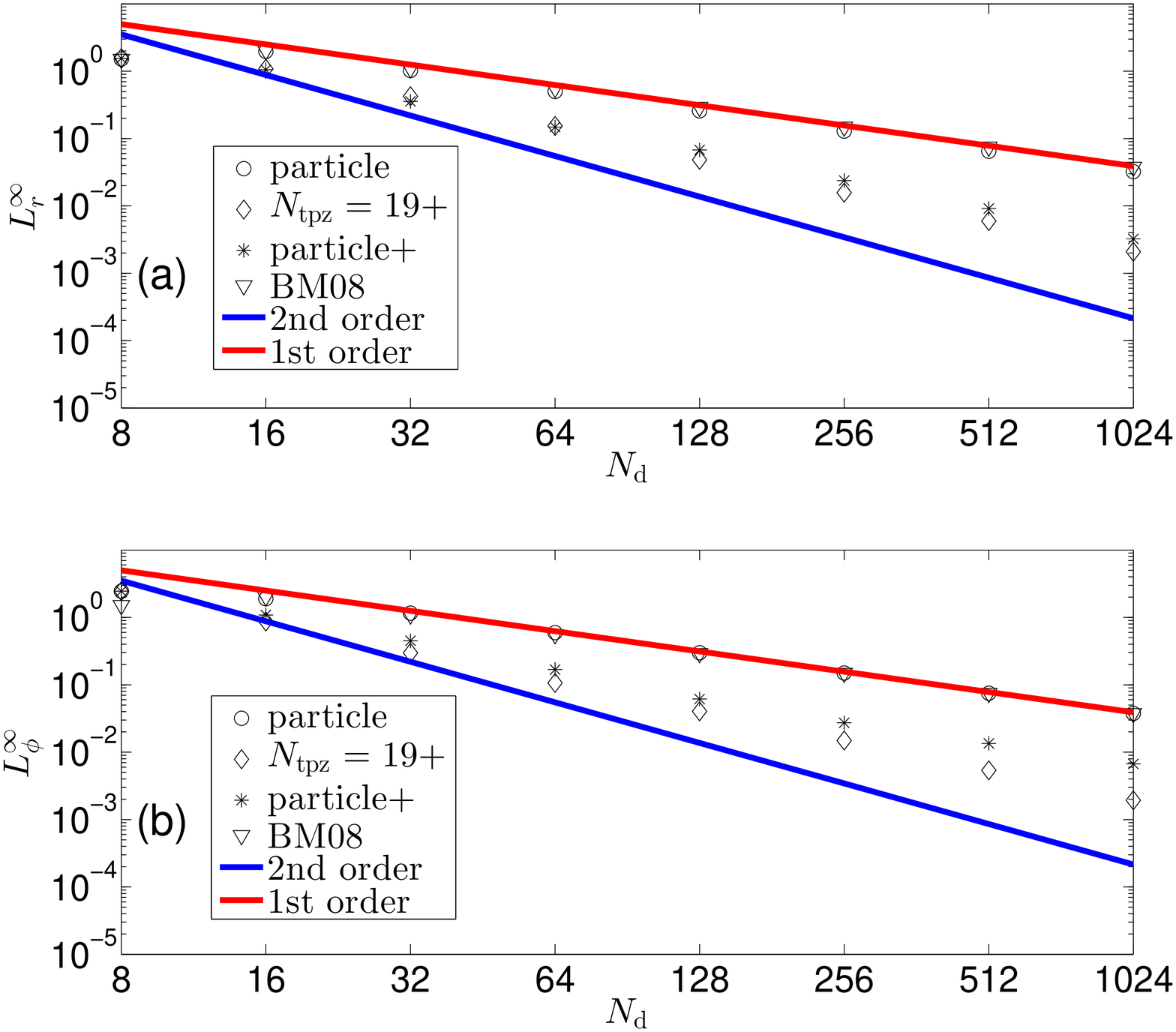}
\caption{The maximum norm errors as a function of cell number $N_{\rm d}$ for the $\sigma_{D_2}$ model with $\alpha=0.25$ in cylindrical coordinates. The center of $\sigma_{D_2}$ is placed at $(x_c=0.5, y_c=0.1)$. The maximum errors of the radial forces are shown in (a) and that of the azimuthal forces are shown in (b). The open circles and asterisks are obtained from the particle-based method without and with density slope correction, respectively. The diamonds are from the SIM proposed in Section~4 with $N_{\rm tpz}=19$. The inverse triangles are obtained using the method described in \citetalias{Bar08}. The red and blue lines indicate the slope of 1st-order and 2nd-order convergence, respectively. \label{fig:Dn2_order_acc}}
\end{figure}

\begin{figure}
\epsscale{1}
\plotone{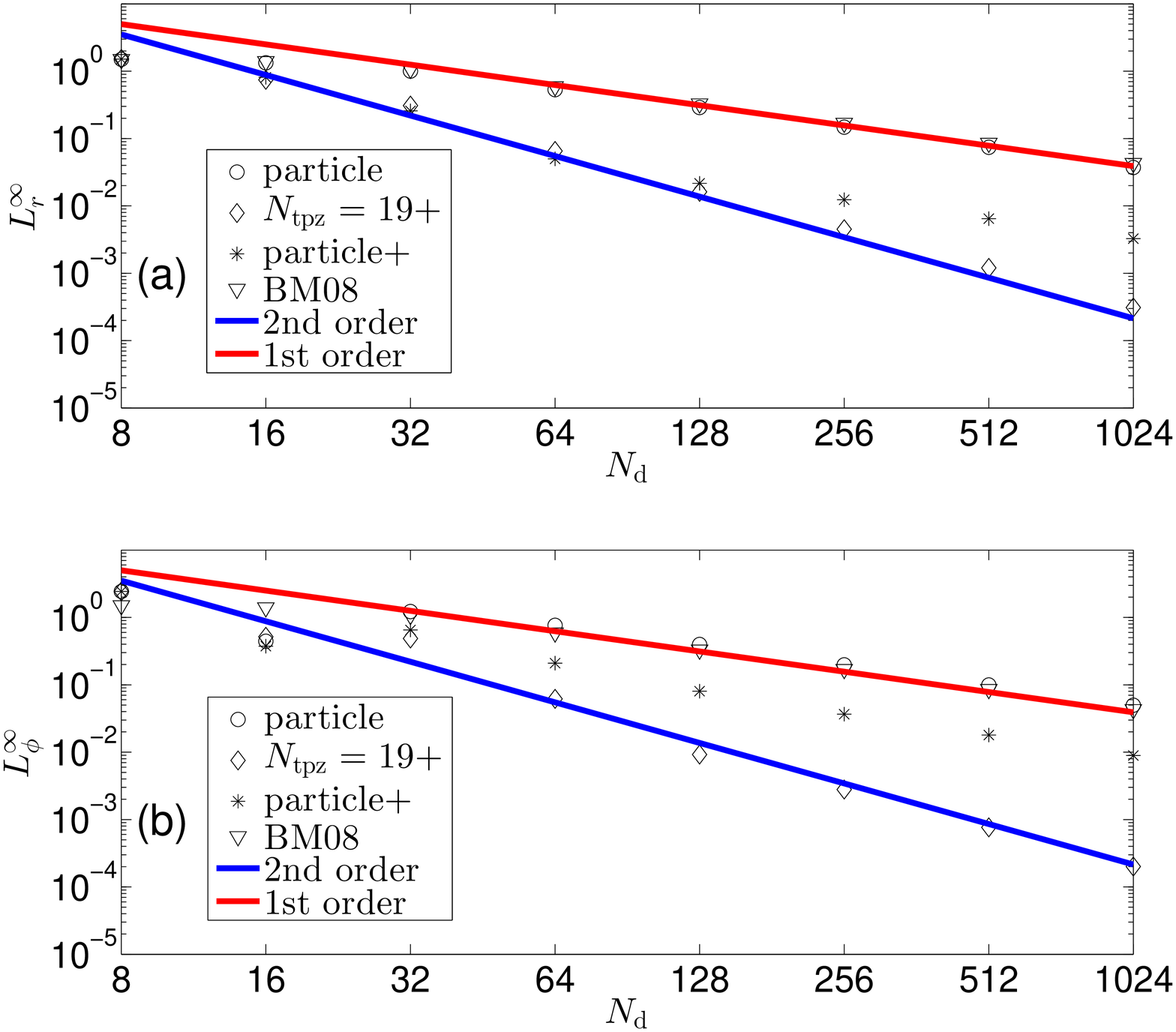}
\caption{The maximum norm errors as a function of cell number $N_{\rm d}$ for the $\sigma_{D_5}$ model with $\alpha=0.25$ in cylindrical coordinates. The center of $\sigma_{D_5}$ is placed at $(x_c=0.5, y_c=0.1)$. The maximum errors of the radial forces are shown in (a) and that of the azimuthal forces are shown in (b). The open circles and asterisks are obtained from the particle-based method without and with density slope correction, respectively. The diamonds are from the SIM proposed in Section~4 with $N_{\rm tpz}=19$. The inverse triangles are obtained using the method described in \citetalias{Bar08}. The red and blue lines indicate the slope of 1st-order and 2nd-order convergence, respectively. \label{fig:Dn5_order_acc}}
\end{figure}

In Figures~\ref{fig:Dn2_order_acc} and \ref{fig:Dn5_order_acc} we show the maximum norm errors for the $\sigma_{D_2}$ disk and the $\sigma_{D_5}$ disk, which are centered at $(x_c=0.5, y_c=0.1)$ and with $\alpha=0.25$. In contrast from the first case, the center of the 
disk is shifted from the origin of cylindrical coordinates, providing nonzero azimuthal forces in 
the domain $\mathcal{R}^{\rm d}$. Both cases involves the integration around singularities and is therefore 
useful for testing the algorithm discussed in Section~4. Since $L^{\infty}$ is particularly helpful for monitoring the convergence of the maximum error in the computational domain, only $L^{\infty}$ is shown in the form of figure. The order of convergence of different algorithms associated with $L^{1}$, $L^{2}$ and $L^{\infty}$ are tabulated in Table~\ref{table:err_order} for both $\sigma_{D_2}$ and $\sigma_{D_5}$ disks.

The top panel of Figures~\ref{fig:Dn2_order_acc} and \ref{fig:Dn5_order_acc} shows the maximum norm errors of the radial forces, while the bottom panel shows that of the azimuthal forces. Four different numerical algorithms are shown in the plots. The open circles are the results obtained from the particle method, diamonds are from the SIM described in Section~4 using $N_{\rm tpz}=19$, the asterisks are from the modified particle method and the triangles are obtained from the algorithms described in \citetalias{Bar08}. When implementing the last algorithm, a softening length $\epsilon=0.015r$ is adopted as that was used in \citetalias{Bar08} \hhw{for the model with $N_d=64$. For a fair comparison, the size of softening length used for other models is scaled linearly with the mesh size. For instance, the softening length used for $N_d=128$ is $\epsilon=7.5\times 10^{-3}r$, while for $N_d=32$ is $\epsilon=0.03r$. We note that the use of a softening length in \citetalias{Bar08} is well} \rt{guided} \hhw{by the physical consideration of disk thickness \citep{Mul12}. These plots and Table~\ref{table:err_order} show that both the particle method (denoted as particle) and the method used in BM08 have genuine first order of accuracy, since the error decreases linearly with the cell size.} Compared to the results of the particle method \hhw{and BM08}, the modified particle method (denoted as particle+) effectively reduces the numerical errors. The difference between the particle, \hhw{the BM08} and the modified particle methods is only on the treatment of a singularity involved in Equations~(30) to (32). Although the improvement on the numerical accuracy is significant, the modified particle method is still of first order convergence. Another significant improvement is achieved when the SIM described in Section~4 is adopted. In addition to the treatment of the singularity, numerical integration of Equations~(30)-(32) is improved using the trapezoidal rule with more than one trapezoid. In this work, we find that using $N_{\rm tpz}>5$ gives reasonable numerical accuracy and the results will not significantly change with increasing $N_{\rm tpz}$. In the case with $\sigma_{D_5}$, the order of convergence is nearly second order. For the model $\sigma_{D_5}$ with $N_{\rm d}=1024$, SIM is two order of magnitude more accurate than the particle \hhw{and the BM08 methods}, and one order of magnitude more accurate than the modified particle method. Meanwhile, the computational complexities of all these four methods are the same $O(N^2\log_2 N)$. 

\begin{figure}
\epsscale{0.9}
\plotone{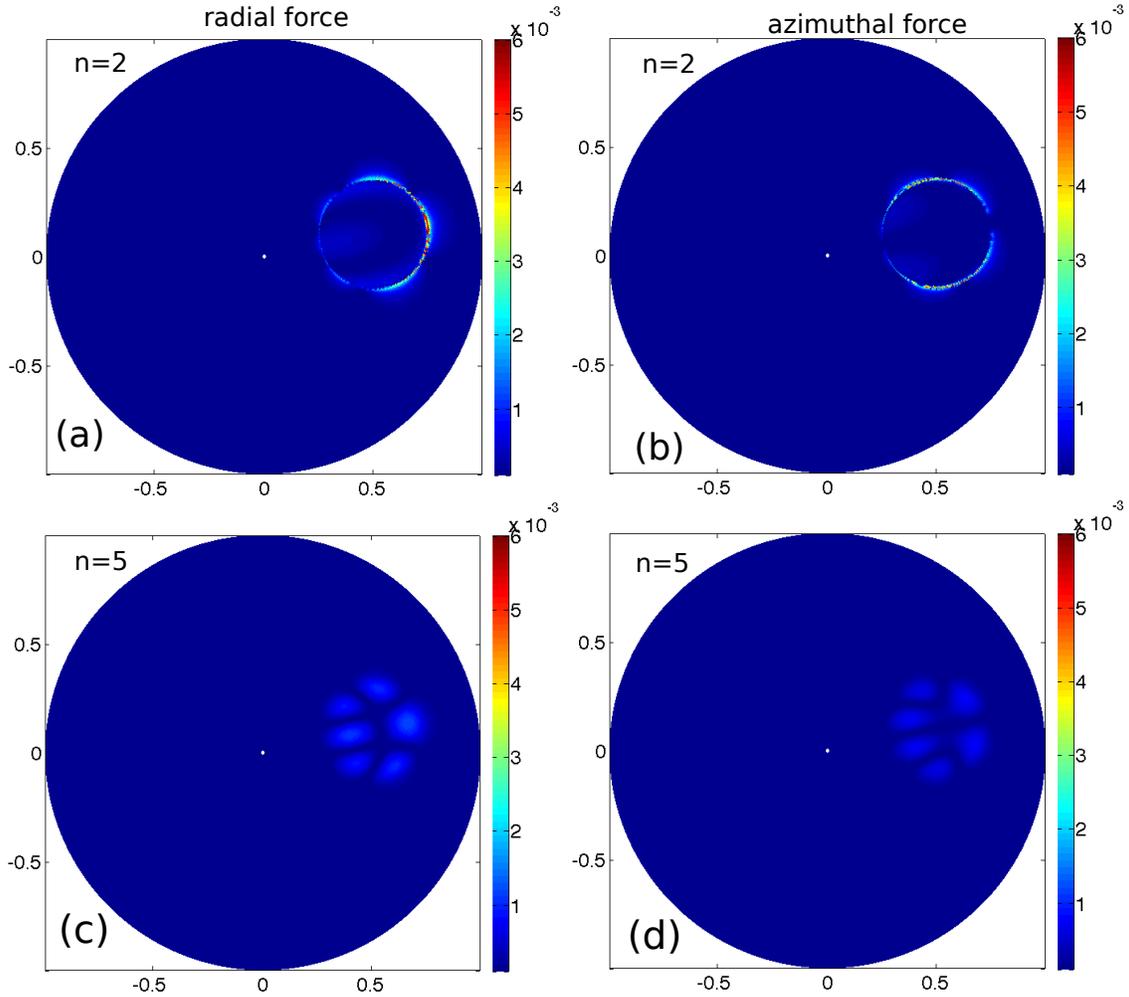}
\caption{The maps of absolute error for the $\sigma_{D_2}$ and $\sigma_{D_5}$ disks discussed in Figures~\ref{fig:Dn2_order_acc} and \ref{fig:Dn5_order_acc}. These maps are obtained using SIM proposed in Section~4. The left column is the results for the radial forces, while the right column is for the azimuthal forces. The top panel is for the $\sigma_{D_2}$ disk, while the bottom one is for the $\sigma_{D_5}$ disk. \label{fig:err_map}}
\end{figure}

From Figure~\ref{fig:Dn2_order_acc} and Table~\ref{table:err_order}, for the $\sigma_{D_2}$ disk, SIM seems to have only roughly 1.5 order of convergence in terms of $L^{\infty}$ while it has nearly 2nd-order accuracy in terms of global error measurement using $L^{1}$ and $L^{2}$. This indicates that the maximum error in $\sigma_{D_2}$ disk converges slower than that in $\sigma_{D_5}$. In Figure~\ref{fig:err_map}, the maps of absolute error are used to investigate the distribution of error for $\sigma_{D_2}$ and $\sigma_{D_5}$ disks. These maps are produced using the SIM. The values of error are color coded as indicated by the corresponding color bars. It is evident that the major errors are concentrated at the edge of the $\sigma_{D_2}$ disk. On the other hand, the errors of the $\sigma_{D_5}$ disk are smoothly distributed over the disk. The reason is that the $\sigma_{D_2}$ disk described by Equation~(50) cannot be well approximated by the linear expansion Equation~(25) when approaching the edge of the 
disk. That is, for $n=2$, a singularity develops in the second derivative of Equation~(50) at $r=\alpha$. This phenomenon does not occur in the $\sigma_{D_5}$ disk since it has a smooth second derivative throughout 
the computational domain. Figure~\ref{fig:err_map} justifies the third requirement for the disk model as mentioned at the beginning of Section~5.1, i.e., the density distribution of the model disk needs to be 
sufficiently smooth for analyzing the order of accuracy.  

\section{Discussion and summary}
Equation~(62) indicates that the particle method is of nearly 2nd-order convergence in cylindrical 
coordinates when $|\vect{x}-\vect{x}'| \gg \Delta r_{i'}$. Similar argument and conclusion can be also 
applied to Cartesian coordinates. This is a desirable property for the calculation of gravitational 
interaction between patches if their domains are mutually exclusive and separated. This situation is 
commonly seen in a numerical code featured with adaptive mesh refinement. We suggest to use the particle 
method for the gravitational interaction between two separated patches, and apply the SIM only for the 
self-gravitational forces inside a patch. In the following, we summarize this work. 

Building on the work of \citet{Yen12}, the self-gravitational force calculation for an 
infinitesimally thin gaseous disk in cylindrical coordinates is improved. The original method proposed 
for the cylindrical coordinates \citep{Yen12} is only a scheme of approximate first order in 
convergence. We identify two sources of error that degrade the numerical accuracy and the order of 
convergence. One arises from the use of the trapezoidal rule for the kernel integration with 
only one trapezoid, the other is due to the presence of singularity in the kernel integral when 
$\vect{x}'=\vect{x}$. The former issue is resolved by increasing the number of trapezoids for 
the integration, while for the latter we adopt the 2nd-order method in Cartesian coordinates, which is 
free of singularity, to evaluate the integral around a singularity in polar coordinates. We prove 
that the result of integration obtained for a specific cell can be applied to other cells if the radial 
direction is discretized logarithmically and the azimuthal direction is discretized evenly. These two 
improvements significantly reduce the numerical error and result in nearly 2nd order convergence.

A similar consideration is applied to the particle method. We show that the particle method is of 
2nd-order convergence in the absence of a singularity. When singularities are involved in 
the self-gravitational calculation, the use of softening length degrades the numerical accuracy. Thus, 
we propose to incorporate the force integration around a singularity as for the SIM to 
the particle method. As a result, the accuracy is significantly improved for $\sigma_{D_2}$ and $\sigma_{D_5}$ disks. However, this correction is not sufficient for improving the order of 
convergence since neglecting the detailed distribution of mass in the surroundings of a singularity 
introduces an error of first order. 

These considerations do not increase the computational complexity $O(N^2\log_{2}N)$ since all the efforts 
are focused on improving the accuracy of force kernels, which are only calculated once at the beginning 
of the simulations.  The method for self-gravitational forces presented here can be similarly applied to 
the gravitational potential~\citep{Yen14}.  

\hhw{Figures~\ref{fig:Dn2_order_acc} and \ref{fig:Dn5_order_acc} show} \rt{a} \hhw{different rate of convergence 
in terms of $L^{\infty}$ for} \rt{the} \hhw{$\sigma_{D_2}$ and $\sigma_{D_5}$ disk, respectively. By noting that the maximum errors are concentrated at the edge of the disk as shown in Figure~\ref{fig:err_map}, we conclude that the rate of convergence is related to the smoothness of} \rt{the} \hhw{mass distribution. We note that the second derivative of $\sigma_{D_2}$ disk does not exist at the edge of disk. This can be also understood mathematically from Equations~(23) and (24). The integrations involve two parts, one is associated with the force kernels and the other is associated with the surface density. Numerical experiment shows that the results for $\sigma_{D_2}$ disk do not significantly change as the number of $N_{\rm tpz}$ increases from 19 to 39. This indicates that an improvement on the force kernel integral by increasing the number of $N_{\rm tpz}$ cannot} \rt{increase} \hhw{the numerical accuracy any further. Thus, the} \rt{lack of} 
\hhw{2nd order behavior in terms of $L^{\infty}$ as shown in Figure~\ref{fig:Dn2_order_acc} can only be 
associated with the term related to the approximation of} \rt{the} \hhw{surface density, i.e., Equation~(25). When applying Equation~(25), we implicitly assume the underlying density is sufficiently smooth so that the error of the approximation is of 2nd order. However, this statement is not true at the edge of} \rt{the} 
\hhw{$\sigma_{D_2}$ disk. Thus, we should not expect a 2nd-order accuracy in terms of $L^{\infty}$ for $\sigma_{D_2}$ disk. The smoothness assumption may seem to be a limitation of the SIM. However, given the fact that those density values are given for a set of discretized points, without a priori knowledge about the} \rt{functional form, 
the} \hhw{linear approximation using Equation~(25) is perhaps a good way to reach second order accuracy.} 
\hhw{We note that the smooth regions can be approximated with higher order accuracy, while those regions with discontinuities can be only approximated with lower order accuracy to avoid the Gibbs phenomenon. For instance, a slope limiter is designed to reach second order accuracy in smooth regions and to avoid numerical oscillations around discontinuities when solving hydrodynamic equations with the Godunov method.}
 
We have shown that the use of a softening length \rt{reduces the accuracy of a Poisson solver to} first order 
\citep{Bar08,Li09}. To be commensurate with the second or higher order accuracy of hydrodynamical solvers, the 
use of a softening length \rt{should} be avoided. One limitation of the method described in this work is the 
use of logarithmic radial grid. This grid configuration may be useful if the computational domain \rt{requires 
a large spatial} range, e.g., a protoplanetary disk. \rt{Readers who are interested in 
using a} uniform discretization in the radial grid should refer to the work by \citet{Li09}. 

\acknowledgments
The authors would like to acknowledge the support of the Theoretical Institute for Advanced Research in Astrophysics (TIARA) based in Academia Sinica Institute of Astronomy and Astrophysics (ASIAA). C.~C.~Yen thanks the Institute of Astronomy and Astrophysics, Academia Sinica, Taiwan for their constant support. C.~C.~Yen is supported by Ministry of Science and Technology of Taiwan, under the Grant MOST-103-2115-M-030-003. Thanks to Mr. Sam Tseng for assistance on the computational facilities and resources (TIARA cluster). The authors thank the referee for comments that helped to improve the clarity and presentation of this paper.
\appendix
\section{Full expressions of kernels} 
\subsection{Cartesian coordinates}
The calculation for $F^{y}_{i,j}$ is fully analogous to that for $F^{x}_{i,j}$. With the linear approximation in surface density, $F^{y}_{i,j}$ can be approximated by:
\begin{equation}
F^{y}_{i,j} \approx F^{y,0}_{i,j}+F^{y,y}_{i,j}+F^{y,x}_{i,j}, 
\end{equation}
where 
\begin{eqnarray}
F^{y,0}_{i,j}=\sum^{N_{\rm d}}_{i'=1}\sum^{N_{\rm d}}_{j'=1}\sigma_{i',j'}\mathcal{K}^{y,0}_{i-i',j-j'}, \\
F^{y,y}_{i,j}=\sum^{N_{\rm d}}_{i'=1}\sum^{N_{\rm d}}_{j'=1}\delta^{y}_{i',j'}\mathcal{K}^{y,y}_{i-i',j-j'}, \\
F^{y,x}_{i,j}=\sum^{N_{\rm d}}_{i'=1}\sum^{N_{\rm d}}_{j'=1}\delta^{x}_{i',j'}\mathcal{K}^{y,x}_{i-i',j-j'},
\end{eqnarray}
and 
\begin{eqnarray}
\mathcal{K}^{y,0}_{i-i',j-j'} \equiv \int\int_{D_{i',j'}} \frac{\bar{y}-y_j}{[(\bar{x}-x_i)^2+(\bar{y}-y_j)^2]^{3/2}}{\rm d}\bar{x}{\rm d}\bar{y}, \\
\mathcal{K}^{y,y}_{i-i',j-j'} \equiv \int\int_{D_{i',j'}} \frac{(\bar{y}-y_j)(\bar{y}-y_{j'})}{[(\bar{x}-x_i)^2+(\bar{y}-y_j)^2]^{3/2}}{\rm d}\bar{x}{\rm d}\bar{y}, \\
\mathcal{K}^{y,x}_{i-i',j-j'} \equiv \int\int_{D_{i',j'}} \frac{(\bar{x}-x_{i'})(\bar{y}-y_{j})}{[(\bar{x}-x_i)^2+(\bar{y}-y_j)^2]^{3/2}}{\rm d}\bar{x}{\rm d}\bar{y}. 
\end{eqnarray}
The full expressions of force kernels used in this work can be found in \citet{Yen12} and are summarized as follows for completeness:
\begin{eqnarray}
\mathcal{K}^{x,0}_{i-i',j-j'}&=&-\ln\left( \bar{y}+\sqrt{\bar{x}^2+\bar{y}^2}\right)\left.\right|^{x_u}_{x_l}\left|^{y_u}_{y_l}\right., \\
\mathcal{K}^{y,0}_{i-i',j-j'}&=&-\ln\left( \bar{x}+\sqrt{\bar{x}^2+\bar{y}^2}\right)\left.\right|^{x_u}_{x_l}\left|^{y_u}_{y_l}\right., \\
\mathcal{K}^{x,x}_{i-i',j-j'}&=&(x_i-x_{i'})\mathcal{K}^{x,0}_{i-i',j-j'}+\left(\bar{y}\ln(\bar{x}+\sqrt{\bar{x}^2+\bar{y}^2})\right)\left.\right|^{x_u}_{x_l}\left|^{y_u}_{y_l}\right., \\
\mathcal{K}^{y,y}_{i-i',j-j'}&=&(y_j-y_{j'})\mathcal{K}^{y,0}_{i-i',j-j'}+\left(\bar{x}\ln(\bar{y}+\sqrt{\bar{x}^2+\bar{y}^2})\right)\left.\right|^{x_u}_{x_l}\left|^{y_u}_{y_l}\right., \\
\mathcal{K}^{x,y}_{i-i',j-j'}&=&(y_j-y_{j'})\mathcal{K}^{x,0}_{i-i',j-j'}+\left(-\sqrt{\bar{x}^2+\bar{y}^2}\right)\left.\right|^{x_u}_{x_l}\left|^{y_u}_{y_l}\right., \\
\mathcal{K}^{y,x}_{i-i',j-j'}&=&(x_i-x_{i'})\mathcal{K}^{y,0}_{i-i',j-j'}+\left(-\sqrt{\bar{x}^2+\bar{y}^2}\right)\left.\right|^{x_u}_{x_l}\left|^{y_u}_{y_l}\right.,
\end{eqnarray}
where $x_l=x_{i'-1/2}-x_i$, $x_u=x_{i'+1/2}-x_i$, $y_l=y_{j'-1/2}-y_j$ and $y_u=y_{j'+1/2}-y_j$. 

The corresponding expression of $F^{y}_{i,j}$ used in the particle-based method is the following:
\begin{equation}
F^{y}_{i,j}=F^{y,corr}_{i,j}+\sum^{N_{\rm d}}_{i'=1}\sum^{N_{\rm d}}_{j'=1}\mathcal{K}^{y,p}_{i-i',j-j'}M_{i',j'},
\end{equation}
where
\begin{eqnarray}
\mathcal{K}^{y,p}_{i-i',j-j'}&=&\begin{cases}
                               \dfrac{y_{j'}-y_j}{[(x_{i'}-x_i)^2+(y_{j'}-y_j)^2]^{3/2}}, & i\neq i'\text{ or } j\neq j' \\
                               0, & \text{otherwise}.
                              \end{cases}, \\
F^{y,corr}_{i,j} &\equiv& \sigma_{i,j} \mathcal{K}^{y,0}_{0,0}+\delta^y_{i,j} \mathcal{K}^{y,y}_{0,0}+\delta^x_{i,j}\mathcal{K}^{y,x}_{0,0}.
\end{eqnarray}
Similarly, $\mathcal{K}^{y,0}_{0,0}=\mathcal{K}^{y,x}_{0,0}=0$ due to the odd symmetry with respect to the cell center. 
\subsection{Polar coordinates}
Following Equation~(24):
\begin{equation}
F^{\phi}_{i,j}=\sum^{N_{\rm d}}_{i'=-N_{\rm s}}\sum^{N_{\rm d}}_{j'=1}\int\int_{\mathcal{R}_{i',j'}} \frac{\sigma(\bar{r}_{i'},\bar{\phi}_{j'})\sin(\bar{\phi}-\phi_j)\bar{r}^2}{[\bar{r}^2+r^2_i-2r_i\bar{r}\cos(\bar{\phi}-\phi_j)]^{3/2}}{\rm d}\bar{r}{\rm d}\bar{\phi}. 
\end{equation}
With the linear approximation in the surface density, $F^{\phi}_{i,j}$ with accuracy of 2nd-order can be approximated by:
\begin{equation}
F^{\phi}_{i,j}\approx F^{\phi,0}_{i,j}+F^{\phi,\phi}_{i,j}+F^{\phi,r}_{i,j},
\end{equation}
where 
\begin{eqnarray}
F^{\phi,0}_{i,j}=\sum^{N_{\rm d}}_{i'=1}\sum^{N_{\rm d}}_{j'=1}\sigma_{i',j'}\mathcal{K}^{\phi,0}_{i-i',j-j'}+\sum^0_{i'=-N_{\rm s}+1}\sum^{N_{\rm d}}_{j'=1}\sigma_{i',j'}\mathcal{K}^{\phi,0}_{i-i',j-j'}+\sum^{N_{\rm d}}_{j'=1}\sigma_{-N_{\rm s},j'}\bar{\mathcal{K}}^{\phi,0}_{i+N_{\rm s},j-j'}, \\
F^{\phi,r}_{i,j}=r_i\left[\sum^{N_{\rm d}}_{i'=1}\sum^{N_{\rm d}}_{j'=1}\delta^r_{i',j'}\mathcal{K}^{\phi,r}_{i-i',j-j'}+\sum^0_{i'=-N_{\rm s}+1}\sum^{N_{\rm d}}_{j'=1}\delta^r_{i',j'}\mathcal{K}^{\phi,r}_{i-i',j-j'}+\sum^{N_{\rm d}}_{j'=1}\delta^{r}_{-N_{\rm s},j'}\bar{\mathcal{K}}^{\phi,r}_{i+N_{\rm s},j-j'}\right],\\
F^{\phi,\phi}_{i,j}=\sum^{N_{\rm d}}_{i'=1}\sum^{N_{\rm d}}_{j'=1}\delta^{\phi}_{i',j'}\mathcal{K}^{\phi,\phi}_{i-i',j-j'}+\sum^0_{i'=-N_{\rm s}+1}\sum^{N_{\rm d}}_{j'=1}\delta^{\phi}_{i',j'}\mathcal{K}^{\phi,\phi}_{i-i',j-j'}+\sum^{N_{\rm d}}_{j'=1}\delta^{\phi}_{-N_{\rm s},j'}\bar{\mathcal{K}}^{\phi,\phi}_{i+N_{\rm s},j-j'},
\end{eqnarray}
and
\begin{eqnarray}
\mathcal{K}^{\phi,0}_{i-i',j-j'}&\equiv&\int\int_{\mathcal{R}_{i',j'}}\frac{\bar{r}^2\sin(\bar{\phi}-\phi_j)}{[\bar{r}^2+r_i^2-2\bar{r}r_i\cos(\bar{\phi}-\phi_j)]^{3/2}}{\rm d}\bar{r}{\rm d}\bar{\phi}, \\
\mathcal{K}^{\phi,r}_{i-i',j-j'}&\equiv&\int\int_{\mathcal{R}_{i',j'}}\frac{\bar{r}^2\sin(\bar{\phi}-\phi_j)(\bar{r}-r_{i'})}{r_i[\bar{r}^2+r_i^2-2\bar{r}r_i\cos(\bar{\phi}-\phi_j)]^{3/2}}{\rm d}\bar{r}{\rm d}\bar{\phi},\\
\mathcal{K}^{\phi,\phi}_{i-i',j-j'}&\equiv&\int\int_{\mathcal{R}_{i',j'}}\frac{\bar{r}^2\sin(\bar{\phi}-\phi_j)(\bar{\phi}-\phi_{j'})}{[\bar{r}^2+r_i^2-2\bar{r}r_i\cos(\bar{\phi}-\phi_j)]^{3/2}}{\rm d}\bar{r}{\rm d}\bar{\phi}.
\end{eqnarray}
Introducing some auxiliary symbols:
\begin{eqnarray}
\mathcal{H}^{r}_1 &\equiv & \left\lbrace-\cos(\bar{\phi})\ln\left(-\cos(\bar{\phi})+\frac{\bar{r}}{r_i}+F\left(\frac{\bar{r}}{r_i},\bar{\phi}\right)\right)+\frac{2\cos(\bar{\phi})(\bar{r}/r_i)-1}{F(\bar{r}/r_i,\bar{\phi})}\right\rbrace, \\
\mathcal{H}^{r}_{2} &\equiv & -\left\lbrace(3\cos^2(\bar{\phi})-1)\ln\left(-\cos(\bar{\phi})+\frac{\bar{r}}{r_i}+F\left(\frac{\bar{r}}{r_i},\bar{\phi}\right)\right) \right.\nonumber \\ 
&&\left. + \frac{1}{F(\bar{r}/r_i,\bar{\phi})}\left(-6\frac{\bar{r}}{r_i}\cos^2(\bar{\phi})+3\cos(\bar{\phi})+\frac{\bar{r}^2}{r_i^2}\cos(\bar{\phi})+\frac{\bar{r}}{r_i}\right)\right\rbrace\\
\mathcal{H}^{\phi}_1 &\equiv & -\left\lbrace F\left(\frac{\bar{r}}{r_i},\bar{\phi}\right)+\cos(\bar{\phi})\ln\left(-\cos(\bar{\phi})+\frac{\bar{r}}{r_i}+F\left(\frac{\bar{r}}{r_i},\bar{\phi}\right)\right) \right\rbrace \\
\mathcal{H}^{\phi}_2 &\equiv &-\left\lbrace \left( \frac{\bar{r}}{2r_i}+\frac{3}{2}\cos(\bar{\phi})\right)F\left(\frac{\bar{r}}{r_i},\bar{\phi}\right) \right. \nonumber \\
&&\left. + \left( \frac{3}{2}\cos^2(\phi)-\frac{1}{2}\right)\ln\left(-\cos(\phi)+\frac{\bar{r}}{r_i}+F\left(\frac{\bar{r}}{r_i}, \bar{\phi}\right)\right)\right\rbrace. \\
\mathcal{H}^{\phi}_3 &\equiv &\bar{\phi}\sin(\bar{\phi})\left\lbrace \frac{1}{F(\bar{r}/r_i,\bar{\phi})}\left(-\frac{\bar{r}}{r_i}-\cos(\bar{\phi})+\frac{\bar{r}}{r_i}\cot^2(\bar{\phi})-\cos(\bar{\phi})\cot^2(\bar{\phi})\right) \right. \nonumber \\
&&\left. +\ln\left(-\cos(\bar{\phi})+\frac{\bar{r}}{r_i}+F\left(\frac{\bar{r}}{r_i}, \bar{\phi}\right)\right)\right\rbrace
\end{eqnarray}
The full expressions of force kernels are:

\begin{eqnarray}
\mathcal{K}^{r,0}_{i-i',j-j'}&\approx & \left. \left. -\mathcal{H}^{r}_1(\bar{r},\bar{\phi})\right|^{r_u}_{r_l}\right]^{\phi_u}_{\phi_l}, \\
\mathcal{K}^{r,r}_{i-i',j-j'}&\approx & \left. \left. -\mathcal{H}^{r}_2(\bar{r},\bar{\phi})\right|^{r_u}_{r_l}\right]^{\phi_u}_{\phi_l}-\frac{r_{i'}}{r_i}\mathcal{K}^{r,0}_{i-i',j-j'}, \\
\mathcal{K}^{r,\phi}_{i-i',j-j'}&\approx & \left. \left. -\bar{\phi}\mathcal{H}^{r}_1(\bar{r},\bar{\phi})\right|^{r_u}_{r_l}\right]^{\phi_u}_{\phi_l}+(\phi_j-\phi_{j'})\mathcal{K}^{r,0}_{i-i',j-j'},\\
\mathcal{K}^{\phi,0}_{i-i',j-j'} &= & \left. \left. \mathcal{H}^{\phi}_1(\bar{r},\bar{\phi})\right|^{r_u}_{r_l}\right|^{\phi_u}_{\phi_l}, \\
\mathcal{K}^{\phi,r}_{i-i',j-j'} &= & \left. \left. \mathcal{H}^{\phi}_2(\bar{r},\bar{\phi})\right|^{r_u}_{r_l}\right|^{\phi_u}_{\phi_l}-\frac{r_{i'}}{r_i}\mathcal{K}^{\phi,0}_{i-i',j-j'}, \\
\mathcal{K}^{\phi,\phi}_{i-i',j-j'} &= & \left. \left. \mathcal{H}^{\phi}_3(\bar{r},\bar{\phi})\right|^{r_u}_{r_l}\right]^{\phi_u}_{\phi_l}+(\phi_j-\phi_{j'})\mathcal{K}^{\phi,0}_{i-i',j-j'},
\end{eqnarray} 
where $r_u=r_{i'+1/2}$, $r_l=r_{i'-1/2}$, $\phi_u=\phi_{j'+1/2}-\phi_j$ and $\phi_l=\phi_{j'-1/2}-\phi_j$. 

\section{Derivation of The Recursive Relations}
Without loss of generality, we set $r=1$ to simplify the notation when deriving the recursive relation. The formula for $r\ge \alpha$ in Equation~(53) can be recast as:
\begin{equation}
\alpha^{2n}F^{r,{\rm ana}}_{D_n}=\pi T_1(\alpha)\left(\sum^{n-1}_{k=0}b^n_{2k}T_{2k}(\alpha)\right)\sin^{-1}(\alpha)+\pi T_1(\alpha)\left(\sum^{n-1}_{k=0}a^n_{2k+1}T_{2k+1}(\alpha) \right)\sqrt{1-\alpha^2},
\end{equation}
where $T_1(\alpha)=\alpha$ has been applied. Equation~(52) then reads:
\begin{equation}
F^{r,{\rm ana}}_{n+1}=\frac{(2n+1)\pi}{\alpha^{2n+1}}(\mathcal{F}+\mathcal{G}),
\end{equation}
where
\begin{eqnarray}
\mathcal{F}&=&\sum_{k=0}^{n-1}\int_0^{\alpha}\left(b^n_{2k}T_1 T_{2k}\right)\sin^{-1}(\hat{\alpha}){\rm d}{\hat{\alpha}}, \\
\mathcal{G}&=&\sum_{k=0}^{n-1}\int_0^{\alpha}\left(a^{n}_{2k+1}T_1 T_{2k+1}\right)\sqrt{1-\hat{\alpha}^2}{\rm d}{\hat{\alpha}}.
\end{eqnarray}
Integrate directly for $k=0,1$ in Equation~(B3):
\begin{eqnarray}
\mathcal{F}&=&\frac{b^n_0}{4}(T_1\sqrt{1-\alpha^2}+T_2\sin^{-1}(\alpha))+ \frac{b^n_2}{16}\left(\frac{T_1+T_3}{2}\sqrt{1-\alpha^2}+T_4\sin^{-1}(\alpha)\right)\nonumber \\
&+&\sum^{n-1}_{k=2}\frac{b^n_{2k}}{2}\int_0^{\alpha}(T_{2k+1}+T_{2k-1})\sin^{-1}(\hat{\alpha}){\rm d}{\hat{\alpha}}, 
\end{eqnarray}
where we have used $T_2(\alpha)=2\alpha^2-1$, $T_4(\alpha)=8\alpha^4-8\alpha^2+1$ and the relation $2T_mT_n=T_{m+n}+T_{|m-n|}$. Integration by part for the integral in Equation~(B5) and apply the identity $\int T_n=\frac{1}{2}(\frac{T_{n+1}}{(n+1)}-\frac{T_{n-1}}{(n-1)})$, we have:
\begin{eqnarray}
&&\sum^{n-1}_{k=2}\frac{b^n_{2k}}{2}\int_0^{\alpha}(T_{2k+1}+T_{2k-1})\sin^{-1}(\hat{\alpha}){\rm d}{\hat{\alpha}} \nonumber \\ &=&\sum^{n-1}_{k=2}\frac{b^n_{2k}}{4}\left(\frac{T_{2k+2}}{2k+2}-\frac{T_{2k-2}}{2k-2}\right)\sin^{-1}(\alpha)-\sum^{n-1}_{k=2}\frac{b^n_{2k}}{4}\int^{\alpha}_{0}\frac{\frac{T_{2k+2}}{2k+2}-\frac{T_{2k-2}}{2k-2}}{\sqrt{1-\hat{\alpha}}^2}{\rm d}{\hat{\alpha}} \\
&=&\left[-\frac{b^n_4}{8}T_2-\frac{b^n_6}{16}T_4+\sum^{n}_{k=3}\left(\frac{b^n_{2k-2}-b^{n}_{2k+2}}{8k}\right)T_{2k}\right]\sin^{-1}(\alpha)-\sum^{n-1}_{k=2}\frac{b^n_{2k}}{4}\int^{\alpha}_{0}\frac{\frac{T_{2k+2}}{2k+2}-\frac{T_{2k-2}}{2k-2}}{\sqrt{1-\hat{\alpha}^2}}{\rm d}{\hat{\alpha}} \nonumber \\
\end{eqnarray}
Apply the change of variable $\hat{\alpha}=\cos(\theta)$, where $\theta=\cos^{-1}(\alpha)$, and use the property $T_n(\cos\theta)=\cos(n\theta)$ to the integral in Equation~(B7):
\begin{eqnarray}
&&-\sum^{n-1}_{k=2}\frac{b^n_{2k}}{4}\int^{\alpha}_{0}\frac{\frac{T_{2k+2}}{2k+2}-\frac{T_{2k-2}}{2k-2}}{\sqrt{1-\hat{\alpha}^2}}{\rm d}{\hat{\alpha}} \\
&=&\frac{1}{4}\sum^{n-1}_{k=2}\int^{\cos^{-1}(\alpha)}_{\pi/2}\frac{\cos[(2k+2)\theta]}{2k+2}-\frac{\cos[(2k-2)\theta]}{2k-2}{\rm d}{\theta}\\
&=&\frac{1}{4}\sum^{n-1}_{k=2}b^n_{2k}\left.\left\lbrace\frac{\sin[(2k+2)\theta]}{(2k+2)^2}-\frac{\sin[(2k-2)\theta]}{(2k-2)^2}\right\rbrace\right|^{\cos^{-1}(\alpha)}_{\pi/2} \\
&=&\frac{1}{4}\sum^{n-1}_{k=2}b^n_{2k}\left[\frac{U_{2k+1}}{(2k+2)^2}-\frac{U_{2k-3}}{(2k-2)^2}\right]\sqrt{1-\alpha^2},
\end{eqnarray} 
where $U_n(\alpha)$ is the Chebyshev polynomial of the second kind of order $n$. From Equation~(B10) to Equation~(B11), we have used the relation:
\begin{eqnarray}
\left.\frac{\sin(n\theta)}{n}\right|^{\cos^{-1}(\alpha)}_{\pi/2}&=&\left.\frac{1}{n}\sqrt{1-\cos^2(n\theta)}\right|^{\cos^{-1}(\alpha)}_{\pi/2} \nonumber \\
                       &=&\left.\frac{1}{n}\sqrt{1-T^2_n}\right|^{\alpha}_0 \\
                       &=&\frac{1}{n}\sqrt{1-\alpha^2}U_{n-1}(\alpha).
\end{eqnarray}
The Chebyshev polynomial of the second kind can be expressed using the Chebyshev polynomial of the first kind through:
\begin{eqnarray}
 U_{2k+1}=2\sum^{k}_{j=0}T_{2j+1}, \\
 U_{2k-3}=2\sum^{k-2}_{j=0}T_{2j+1}.
\end{eqnarray}
Therefore, Equation~(B11) can be expressed using the Chebyshev polynomial of the first kind:
\begin{eqnarray}
&&\frac{1}{4}\sum^{n-1}_{k=2}b^n_{2k}\left[\frac{U_{2k+1}}{(2k+2)^2}-\frac{U_{2k-3}}{(2k-2)^2}\right]\sqrt{1-\alpha^2} \\
&=&\frac{\sqrt{1-\alpha^2}}{4}\left\lbrace 2\sum^{n-1}_{k=2}\frac{b^n_{2k}}{(2k+2)^2}\sum^{k}_{j=0}T_{2j+1}-2\sum^{n-1}_{k=3}\frac{b^n_{2k}}{(2k-2)^2}\sum^{k-2}_{j=0}T_{2j+1}-\frac{b^n_4}{2}T_1 \right\rbrace \\
&=& \frac{\sqrt{1-\alpha^2}}{4}\left\lbrace 2\sum^{n-1}_{j=2}\left[\sum^{n-1}_{k=j}\frac{b^n_{2k}}{(2k+2)^2}\right]T_{2j+1}+2\sum^{n-1}_{k=2}\frac{b^n_{2k}}{(2k+2)^2}(T_1+T_3)-\frac{b^n_4}{2}T_1 \right. \nonumber \\
&&\left. -2\sum^{n-3}_{j=1}\left[\sum^{n-1}_{k=j+2}\frac{b^n_{2k}}{(2k-2)^2}\right]T_{2j+1}-2\sum^{n-1}_{k=3}\frac{b^n_{2k}}{(2k-2)^2}T_1\right\rbrace.
\end{eqnarray}
From Equation~(B17) to Equation~(B18), the order of summation is exchanged. Combining Equations~(B5)(B7)(B11)(B18), $\mathcal{F}$ can be expressed in terms of $\sin^{-1}(\alpha)$, $\sqrt{1-\alpha^2}$ and the combination of Chebyshev polynomials:
\begin{eqnarray}
\mathcal{F}&=&\left[\frac{b^n_0}{8}T_2+\sum^{n}_{k=1}\left(\frac{b^n_{2k-2}-b^n_{2k+2}}{8k}\right)T_{2k}\right]\sin^{-1}(\alpha) \nonumber \\
            &+&\left\lbrace \frac{b^n_0}{8}T_1 +\frac{1}{2}\sum^{n-1}_{j=0}\left[\sum^{n-1}_{k=j}\frac{b^n_{2k}}{(2k+2)^2}\right]T_{2j+1}-\frac{1}{2}\sum^{n-3}_{j=0}\left[\sum^{n-3}_{k=j}\frac{b^n_{2k+4}}{(2k+2)^2}\right]T_{2j+1}\right\rbrace\sqrt{1-\alpha^2}. \nonumber \\
\end{eqnarray}
Similar to the derivation for Equation~(B19),  it is straightforward to show that Equation~(B4) has the following expression:
\begin{eqnarray}
\mathcal{G}&=&\frac{1}{8}(a^n_1-a^n_3)T_0\sin^{-1}(\alpha)+\left[ -\sum^{n-1}_{j=0}\left(\sum^{n-1}_{k=j}\frac{a^n_{2k+1}}{2k+2}\right)T_{2j+1}+\sum^n_{j=0} \left(\sum^n_{k=j}\frac{a^n_{2k-1}}{2k+2}\right)T_{2j+1}\right. \nonumber \\
&-&\left. \sum^{n-2}_{j=0} \left(\sum^{n-2}_{k=j}\frac{a^n_{2k+3}}{2k+2}\right)T_{2j+1}+\sum^{n-3}_{j=0}\left(\sum^{n-3}_{k=j}\frac{a^n_{2k+5}}{2k+2}\right)T_{2j+1}+\frac{a^n_1}{2}T_1\right]\frac{\sqrt{1-\alpha^2}}{4}.
\end{eqnarray}
Substituting Equations~(B19) and (B20) into Equation~(B2), we obtain the recursive relations as shown by Equations~(54) and (55). 

\bibliography{selfgravity}

\begin{thebibliography}{21}
\providecommand\natexlab[1]{#1}
\providecommand\JournalTitle[1]{#1}

\bibitem[{{Baruteau} \& {Masset}(2008)}]{Bar08}
{Baruteau}, C., \& {Masset}, F. 2008,
  \href{http://dx.doi.org/10.1086/529487}{\JournalTitle{\apj}, 678, 483}

\bibitem[{{Binney} \& {Tremaine}(2008)}]{Bin08}
{Binney}, J., \& {Tremaine}, S. 2008, {Galactic Dynamics: Second Edition}
  (Princeton University Press)

\bibitem[{Bracewell(1999)}]{Bra99}
Bracewell, R.~N. 1999, The Fourier Transformation and its Applications, 3rd
  edn. (Mc.~Graw-Hill)

\bibitem[{{Elmegreen} {et~al.}(2014){Elmegreen}, {Elmegreen}, {Erroz-Ferrer},
  {Knapen}, {Teich}, {Popinchalk}, {Athanassoula}, {Bosma}, {Comer{\'o}n},
  {Efremov}, {Gadotti}, {Gil de Paz}, {Hinz}, {Ho}, {Holwerda}, {Kim}, {Laine},
  {Laurikainen}, {Men{\'e}ndez-Delmestre}, {Mizusawa}, {Mu{\~n}oz-Mateos},
  {Regan}, {Salo}, {Seibert}, \& {Sheth}}]{Elm14}
{Elmegreen}, D.~M., {Elmegreen}, B.~G., {Erroz-Ferrer}, S., {et~al.} 2014,
  \href{http://dx.doi.org/10.1088/0004-637X/780/1/32}{\JournalTitle{\apj}, 780,
  32}

\bibitem[{Evans(1991)}]{Eva91}
Evans, L.~C. 1991, Graduate Studies in Mathematics, Vol.~19, {Partial
  Differential Equations} (Providence, Rhode Island)

\bibitem[{{Hockney} \& {Eastwood}(1988)}]{Hoc88}
{Hockney}, R.~W., \& {Eastwood}, J.~W. 1988, {Computer simulation using
  particles}

\bibitem[{{Inutsuka} {et~al.}(2010){Inutsuka}, {Machida}, \&
  {Matsumoto}}]{Inu10}
{Inutsuka}, S.-i., {Machida}, M.~N., \& {Matsumoto}, T. 2010,
  \href{http://dx.doi.org/10.1088/2041-8205/718/2/L58}{\JournalTitle{\apjl},
  718, L58}

\bibitem[{James(1977)}]{Jam77}
James, R.~A. 1977,
  \href{http://dx.doi.org/http://dx.doi.org/10.1016/0021-9991(77)90013-4}{\JournalTitle{JCoPh},
  25, 71}

\bibitem[{{Kim} {et~al.}(2012){Kim}, {Seo}, \& {Kim}}]{Kim12}
{Kim}, W.-T., {Seo}, W.-Y., \& {Kim}, Y. 2012,
  \href{http://dx.doi.org/10.1088/0004-637X/758/1/14}{\JournalTitle{\apj}, 758,
  14}

\bibitem[{{Lee}(2014)}]{Lee14}
{Lee}, W.-K. 2014,
  \href{http://dx.doi.org/10.1088/0004-637X/792/2/122}{\JournalTitle{\apj},
  792, 122}

\bibitem[{{Lee} \& {Shu}(2012)}]{Lee12}
{Lee}, W.-K., \& {Shu}, F.~H. 2012,
  \href{http://dx.doi.org/10.1088/0004-637X/756/1/45}{\JournalTitle{\apj}, 756,
  45}

\bibitem[{{Li} {et~al.}(2009){Li}, {Buoni}, \& {Li}}]{Li09}
{Li}, S., {Buoni}, M.~J., \& {Li}, H. 2009,
  \href{http://dx.doi.org/10.1088/0067-0049/181/1/244}{\JournalTitle{\apjs},
  181, 244}

\bibitem[{{Lin} {et~al.}(2013){Lin}, {Wang}, {Hsieh}, {Taam}, {Yang}, \&
  {Yen}}]{Lin13}
{Lin}, L.-H., {Wang}, H.-H., {Hsieh}, P.-Y., {et~al.} 2013,
  \href{http://dx.doi.org/10.1088/0004-637X/771/1/8}{\JournalTitle{\apj}, 771,
  8}

\bibitem[{{Mestel}(1963)}]{Mes63}
{Mestel}, L. 1963, \JournalTitle{\mnras}, 126, 553

\bibitem[{{M{\"u}ller} {et~al.}(2012){M{\"u}ller}, {Kley}, \& {Meru}}]{Mul12}
{M{\"u}ller}, T.~W.~A., {Kley}, W., \& {Meru}, F. 2012,
  \href{http://dx.doi.org/10.1051/0004-6361/201118737}{\JournalTitle{\aap},
  541, A123}

\bibitem[{{Schulz}(2009)}]{Sch09}
{Schulz}, E. 2009,
  \href{http://dx.doi.org/10.1088/0004-637X/693/2/1310}{\JournalTitle{\apj},
  693, 1310}

\bibitem[{{Seo} \& {Kim}(2014)}]{Seo14}
{Seo}, W.-Y., \& {Kim}, W.-T. 2014,
  \href{http://dx.doi.org/10.1088/0004-637X/792/1/47}{\JournalTitle{\apj}, 792,
  47}

\bibitem[{{Yen}(2014)}]{Yen14}
{Yen}, C.-C. 2014,
  \href{http://dx.doi.org/10.1155/2014/693537}{\JournalTitle{SJAM}, 2014}

\bibitem[{{Yen} {et~al.}(2012){Yen}, {Taam}, {Yeh}, \& {Jea}}]{Yen12}
{Yen}, C.-C., {Taam}, R.~E., {Yeh}, K.~H.-C., \& {Jea}, K.~C. 2012,
  \href{http://dx.doi.org/10.1016/j.jcp.2012.08.003}{\JournalTitle{JCoPh}, 231,
  8246}

\bibitem[{{Zhang} {et~al.}(2014){Zhang}, {Liu}, {Zhou}, \&
  {Wittenmyer}}]{Zha14}
{Zhang}, H., {Liu}, H.-G., {Zhou}, J.-L., \& {Wittenmyer}, R.~A. 2014,
  \href{http://dx.doi.org/10.1088/1674-4527/14/4/006}{\JournalTitle{RAA}, 14,
  433}

\bibitem[{{Zhang} {et~al.}(2008){Zhang}, {Yuan}, {Lin}, \& {Yen}}]{Zha08}
{Zhang}, H., {Yuan}, C., {Lin}, D.~N.~C., \& {Yen}, D.~C.~C. 2008,
  \href{http://dx.doi.org/10.1086/528707}{\JournalTitle{\apj}, 676, 639}

\end{thebibliography}
\clearpage

\begin{table}[tp]%
\caption{Order of convergence in terms of $L^1$, $L^2$ and $L^{\infty}$ for the radial and azimuthal forces for those algorithms discussed in Figures~\ref{fig:Dn2_order_acc} and \ref{fig:Dn5_order_acc}. The numbers are extracted using the numerical errors obtained for $N_d=64, 128, 256, 512, 1024$.}
\label{table:err_order} \centering
\begin{tabular}{cccccc}
\toprule
disk model 		&  error norm		& particle		& SIM		& particle+		& BM08 \\
\midrule
$\sigma_{D_2}$ 	&$L^{\infty}_r$		& 1.0 			& 1.5 		& 1.4 			& 1.0 \\
               	&$L^{\infty}_{\phi}$	& 1.0			& 1.4		& 1.1			& 1.0 \\
               	&$L^2_r$				& 1.0			& 1.9		& 1.2			& 1.0 \\
               	&$L^2_{\phi}$		& 1.0			& 2.0		& 1.1			& 1.0 \\
               	&$L^1_r$				& 1.0			& 2.0		& 1.2			& 1.0 \\
               	&$L^1_{\phi}$		& 1.0			& 2.1		& 1.2			& 1.0 \\
\midrule
$\sigma_{D_5}$	&$L^{\infty}_r$		& 1.0			& 1.9		& 1.0			& 1.0 \\
				&$L^{\infty}_{\phi}$	& 1.0			& 2.0		& 1.1			& 1.0 \\
				&$L^2_r$				& 1.0			& 1.9		& 1.0			& 1.0 \\
				&$L^2_{\phi}$		& 1.0			& 1.9		& 1.1			& 1.0 \\
				&$L^1_r$				& 1.0			& 1.9		& 1.1			& 1.0 \\
				&$L^1_{\phi}$		& 1.0			& 1.9		& 1.1			& 1.0 \\
\bottomrule
\end{tabular}
\end{table}

\end{document}